\documentclass[11pt]{article}
\usepackage{amsthm,amsmath,amsfonts,amssymb,dsfont, bbm}
\usepackage{graphicx,psfrag,epsf}
\usepackage{enumerate}
\usepackage{natbib}
\usepackage{url} % not crucial - just used below for the URL 
\usepackage[margin=1in]{geometry}
\usepackage{amsfonts}
\usepackage{algorithm}
\usepackage{algpseudocode}
\usepackage{dsfont}
\usepackage[T1]{fontenc}

\newtheorem{assumption}{Assumption}
\usepackage[font=footnotesize]{caption}

\usepackage{booktabs}

\newcommand{\E}{\mathbb{E}}

\newcommand{\R}{{\mathds R}}
\newcommand{\x}{{x}}
\newcommand{\X}{{X}}

\newcommand{\Norm}{\mathcal{N}}
\newcommand{\PM}{\textsc{pm}$_{2.5}$ }
\newcommand{\PMns}{\textsc{pm}$_{2.5}$}

\newcommand{\indep}{\perp \!\!\! \perp}

\usepackage[table, dvipsnames]{xcolor} 
\usepackage[resetlabels]{multibib}
\newcommand{\sigmat}{{\sigma^{(t)}}}
\newcommand{\betat}{\beta^{(t)}}
\newcites{App}{References}
\usepackage{hyperref}
\hypersetup{
colorlinks=true,
citecolor=blue,
linkcolor=black,
filecolor=blue,      
urlcolor=black,
}
\definecolor{lightGray}{HTML}{DEDEDE}
\definecolor{Gray2}{HTML}{F7F7F7}

\begin{document}

\def\spacingset#1{\renewcommand{\baselinestretch}%
{#1}\small\normalsize} \spacingset{1}

%%%%%%%%%%%%%%%%%%%%%%%%%%%%%%%%%%%%%%%%%%%%%%%%%%%%%%%%%%%%%%%%%%%%%%%%%%%%%%

{
  \title{\LARGE \bf  Characterizing the Effects of Environmental Exposures on Social Mobility: Bayesian Semi-parametrics for Principal Stratification
  }
   \author{Dafne Zorzetto$^*$\\
 Data Science Institute, Brown University, Rhode Island, USA \vspace{.2cm}\\ 
 Paolo Dalla Torre$^*$ \\
 Department of Decision Sciences, Bocconi University, Italy \vspace{.2cm}\\
 Sonia Petrone \\
 Department of Decision Sciences, Bocconi University, Italy \vspace{.2cm}\\
 Francesca Dominici \vspace{.1cm}\\ 
 Department of Biostatistics, Harvard School of Public Health, Massachusetts, USA \vspace{.3cm}\\ 
 Falco J. Bargagli-Stoffi \vspace{.1cm}\\
Department of Biostatistics, University of California, Los Angeles\\
\texttt{falco@g.ucla.edu}}

\date{} 
\maketitle

\begin{abstract}
Understanding the causal effects of air pollution exposures on social mobility is attracting increasing attention. At the same time, education is widely recognized as a key driver of social mobility. However, the causal pathways linking fine particulate matter (\PMns) exposure, educational attainment, and social mobility remain largely unexplored. To address this, we adopt the principal stratification approach, which rigorously defines causal effects when a post-treatment variable---i.e., educational attainment---is affected by exposure---i.e., \PMns---and may, in turn, affect the primary outcome---i.e., social mobility. To estimate the causal effects, we propose a Bayesian semi-parametric method leveraging infinite mixtures for modeling the primary outcome. The proposed method (i) allows flexible modeling of the distribution of the primary potential outcomes, (ii) improves the accuracy of counterfactual imputation---a fundamental problem in causal inference framework---, and (iii) enables the characterization of treatment effects across different values of the post-treatment variable. We evaluate the performance of the proposed methodology through a Monte Carlo simulation study, demonstrating its advantages over existing approaches. Finally, we apply our method to a national dataset of 3,009 counties in the United States to estimate the causal effect of \PM on social mobility, taking into account educational attainment as a post-treatment variable. Our findings indicate that in counties where higher \PM exposure significantly reduces educational attainment social mobility decreases by approximately 5\% compared to counties with lower \PM exposure. We also find that in counties where exposure to \PM does not affect educational attainment, social mobility is reduced by approximately 2\% hinting at the possibility of further, yet unexplored, pathways connecting air pollution and social mobility.
\end{abstract}

\noindent%
{\it Bayesian Semi-parametric, Causal Inference, Education, Principal Stratification, Posterior Predictive Distribution}
\vfill

%\newpage
\spacingset{1.9} % DON'T change the spacing!

%\listoftodos

\large

\section{Introduction}
\label{sec: intro}

\subsection{Motivation}
\label{subsec:motivation}
The adverse effects of fine particulate matter (\PMns) on human health are well established \citep{Dominici2000,dominici2014particulate, dominici2022assessing}. Yet, its impact on socio-economic indicators remains less explored and understood. Recent studies investigate the association between \PM and \textit{social mobility}. Social mobility is a key socio-economic indicator that measures the extent to which individuals or families can improve their economic and social status over time relative to previous generations \citep{Obrien2018, manduca2021childhood, swetschinski2023exposures}. Social mobility is important because it reflects the fairness and opportunity structure of society, serving as a fundamental measure of whether economic prosperity and advancement remain accessible across different social strata and geographic communities.

The seminal work of \citet{Lee2024} found a negative effect of \PM exposure on social mobility. This indicates that exposure to higher level of \PM induces a significant reduction in the economic and social status of the next generation compared to the previous one. 

Education plays a crucial role in social mobility \citep{chetty2014land}. The causal analysis by \citep{Rauscher2016, Biasi2023, kratz2022} demonstrates how higher levels of education positively affect an individual's ability to improve social mobility. Simultaneously, other studies have underscored the detrimental effects of \PM exposure on educational attainment \citep{sunyer2015association,currie2009air}, suggesting an intersection between environmental factors and educational outcomes. 

In light of the above, our aim is to answer the question \textit{What is the causal effect of \PM exposure on social-mobility across different levels of educational attainments?}  To answer this question, we turn to the principal stratification approach \citep{frangakis2002principal}. In fact, this approach provides a means for estimating the causal effect of a treatment---in our case, the \PM exposure---on a primary outcome---i.e., social mobility---, while adjusting for the post-treatment variables---i.e., the educational attainment. The causal effect is defined conditionally to the \textit{principal strata}, i.e., groups of units characterized by the joint values of the potential outcome of the post-treatment variable \citep{antonelli2023principal, mealli2012refreshing}.

In principal stratification, the \textit{dissociative effect} captures the causal effect of treatment on the primary outcome for the stratum composed of units that have no effect of treatment on the post-treatment variable. In the case of our motivating application, this would represent the effect of exposure to air pollution on social mobility among the units where the is no evidence that the exposure to \PM does not affect the level of education. Conversely, the \textit{associative negative (positive) effect} represents the causal effect of treatment on the outcome for the principal stratum composed of units whose post-treatment variable is decreased (increased) by treatment \citep{frangakis2002principal, mealli2012refreshing}. In our application, we are interested in quantifying the causal effect of air pollution exposure on social mobility in those units in which the education level is positively or negatively affected by the exposure to \PMns.

\subsection{Methodological Contributions}
\label{subsec:contribution}

In this paper, we define a novel Bayesian approach for principal stratification where we consider a binary treatment---indicating a high or low level of \PM exposure with respect to the empirical median---, a continuous post-treatment variable---the educational attainment defined as the percentage of graduation rate from high school, community college or college---and a continuous outcome---the social mobility measure. Although existing work on principal stratification has focused mainly on binary or discrete post-treatment variables \citep[e.g.,][]{angrist1996identification, imai2008sharp, ding2011identifiability, mattei2011augmented, mealli2013using, mealli2016identification, jiang2022multiply, mattei2024assessing, ohnishi2024bayesian, gravelle2024bayesian, sisti2024bayesian}, the framework to accommodate a continuous post-treatment variable, as in our setting, requires a more refined formulation of the causal estimand and a flexible modeling definition \citep{jin2008principal, conlon2014surrogacy, lu2023principal, schwartz2011bayesian, sun2024principal}.

We define a novel Bayesian semi-parametric model in which the potential post-treatment variable, conditional to the confounders, is modeled using a Gaussian linear regression, while the potential primary outcome, conditional to confounders and potential post-treatment variable, is modeled with an infinite mixture distribution through a Bayesian nonparametric (BNP) prior.

BNP methods have recently received increasing attention in the causal inference literature for their ability to flexibly model complex distributions while accurately quantifying uncertainty \citep{linero2021and}. This has led to a growing body of work employing Gaussian processes \citep{branson2019nonparametric, ray2019debiased, vegetabile2020optimally} and the Dirichlet process \citep{kim2017framework, roy2018bayesian, oganisian2021bayesian, zorzetto2024confounder, hu2023bayesian}. Within the principal stratification framework, \cite{antonelli2023principal} define BNP models for the potential outcome of the continuous post-treatment variable under continuous treatment variable; \cite{schwartz2011bayesian} define a Dirichlet process mixture for the bivariate distribution of the potential post-treatment variables under a binary treatment;
while \cite{zorzetto2024confounder} introduce a dependent Dirichlet process with hierarchical structure across treatment levels for the continuous potential post-treatment variables.

However, although these principal stratification approaches employ flexible BNP models for the post-treatment variable, they define a parametric specification for the outcome distribution, assuming a linear relationship with both the covariates and the post-treatment variable. A notable exception is \citet{antonelli2023principal}, although their work focus on a continuous treatment variable.  

In our work, we relax the parametric modeling assumptions for the primary outcome by leveraging a dependent Dirichlet process \citep[for a review refer to][]{mac2000dependent, quintana2020dependent}.  
This approach defines the primary outcome distribution as a Dirichlet process mixture that flexibly depends on both the covariates and the post-treatment variable. Specifically, our method (i) allows a more flexible distribution for potential primary outcomes compared to the literature, consequently, (ii) improves the accuracy of data imputation of the counterfactual outcome---a missing data problem that is fundamental in the causal inference framework---, and (iii) allows the characterization of treatment effects across different values of potential post-treatment variable.

\subsection{Organization of the Article} 

Section \ref{sec:motivation} introduces the motivating application and the questions that we aim to address. In Section \ref{sec:set_up}, we introduce the notation, assumptions that we use throughout the paper, and the principal causal effects. In Section \ref{sec:model} we define the Bayesian semi-parametric model. The simulation study to assess the performance of our proposed model in different scenarios is reported in Section \ref{sec:simulations}. The description of the data set and the results of the application of are in Section \ref{sec:application}. Section \ref{sec:conclusion} concludes the paper with a discussion of the proposed model and further research.
\section{Motivating Application}
\label{sec:motivation}

\subsection{Social Mobility's Factors} 
Recent empirical evidence suggests a concerning trend in US social mobility, with studies documenting stagnation or decline in intergenerational economic mobility \citep{piketty2018distributional, song2019firming}. A seminal study by \citet{chetty2017fading} found that the percentage of children who earn more than their parents has decreased from approximately 90\% for children born in 1940 to 50\% for children born in the 1980s.

The decline in intergenerational mobility raises significant social concerns, as reduced economic mobility can weaken social cohesion. In fact, as argued by \citet{stiglitz2012price}, the increase in inequality often equates to a decrease in equal opportunity and exacerbating economic and social inequalities. Recent literature has focused on searching for the root causes of the decline in economic mobility. While multiple factors---such as cultural \citep{platt2019understanding}, demographic \citep{van2011family, salvanes2023drives}, geographical \citep{connor2020changing, salvanes2023drives}, labor market \citep{choi2023social}, welfare state \citep{heckman2022lessons} as well as macro-economic trends \citep{piketty2018distributional}---play an important role, recent work have highlighted that \textit{environmental} and \textit{educational factors} might be key determinants of the decline in social mobility.

Specifically, in a recent contribution, \citet{Lee2024} identified a direct causal link between \PM exposure and social mobility, finding that a $1 \mu g/m³$ increase in childhood exposure to fine particulate matter (\PM) leads to a $1.146\%$ decrease in absolute upward mobility. This seminal study highlights the potential for \PM to directly hinder socioeconomic advancement. However, this study falls short of investigating the potential causal pathways through which exposure to higher levels of air pollution hinders social mobility. 

Educational attainment is a key factor in economic and social mobility. Empirical evidence consistently demonstrates its significant returns in terms of earnings and intergenerational mobility. Studies show that higher education increases earnings potential \citep{card2001causal} and enhances intergenerational mobility, helping individuals from lower socio-economic backgrounds improve their status across generations \citep{blanden2014education, pfeffer2015educational, brown2020educational}. From a causal perspective, a few studies have analyzed how educational attainment influences social mobility. \cite{Rauscher2016} found that minimal schooling requirements increased intergenerational mobility in 19th-century U.S., while 20th-century school finance reforms boosted upward mobility for low-income students by narrowing disparities in teacher resources and college access \citep{Biasi2023}. \citet{kratz2022} showed education's dynamic equalizing effects, finding that highly educated individuals experienced a decline in parental influence on occupational status, while low-educated individuals faced a significant increase in direct origin effects.

\subsection{Environmental Exposures and Educational Attainments} 

Evidence suggests that two key socioeconomic factors---environmental exposure and educational attainment---are closely interconnected. As highlighted in \citet{bearer1995children}, children are particularly vulnerable to environmental pollutants, which can compromise health and cognitive development. Exposure during critical developmental periods has been associated with lower IQ, impaired neurological function, and reductions in memory and motor skills. 

Notably, air pollution significantly impacts education outcomes both in the short and long term. Recent studies have revealed that long-term exposure to pollutants earlier in life can lead to decreased academic performance and higher rates of cognitive disabilities and mental health problems \citep{zhang2018impact, braithwaite2019air}. These findings are further supported by studies indicating that even short-term air pollution exposure can impair cognitive function, reduce concentration, and increase absenteeism among students \citep{sunyer2015association, shehab2019effects}. The effects of pollution on education are not limited to individual performance. Air pollution may also contribute to the exacerbation of existing inequalities, since schools in more polluted areas often serve disadvantaged populations \citep{mohai2011air}. Furthermore, there is critically important evidence that suggests that the economic impact of pollution-related educational deficits can be substantial, with significant losses in lifetime earnings for affected students \citep{currie2009air}, which could, in turn, substantially hinder social mobility. 

\subsection{Characterizing the Effects of Environmental Exposures on Social Mobility Across Different Educational Attainments} 

To disentangle the complex causal relationships linking air pollution (as an exposure), education (as a post-treatment), and social mobility (as a primary outcome), we harmonize multiple publicly available datasets and develop a methodology to analyze such data within a coherent principal stratification framework.

\subsection{Study Dataset}\label{subsec:data}

The dataset used in this analysis is constructed from multiple publicly available sources. Specifically, air pollution data on \PM exposure levels were derived from high-resolution satellite estimates at a $0.9km \times 1.1km$ grid scale \citep{colmer2023}, which was then aggregated to the census tract level. Social mobility data, used as the primary outcome, and different levels of education measured as a rate in the considered population, used as post-treatment variables, were taken from the \textit{Opportunity Atlas} \citep{chetty2018opportunity}. The socio-economic and demographic data were obtained from the \textit{U.S. Census} (1990–2000), while meteorological variables were sourced from the \textit{Daymet} dataset (1982) \citep{castro2024daymet}. Both information of socio-economic and demographic information and meteorological data are considered as confounders. These combined sources provided the comprehensive dataset used for our analysis.

The study was conducted at the county level in the continental U.S., encompassing 3,009 counties. Although most variables were initially available at the census tract level, all data was aggregated to the county level to align with education levels, which are only available at this scale.

The final dataset for analysis retains several key variables listed in Table \ref{table: descr stats application dataset}, where their descriptive statistics and data sources are reported. Census data include the percentage of college educated in the year 2000, the median household income in 1990, the population density in the year 2000 per square mile, the share of people who live below poverty levels, and the share of black, white, Hispanic and Asian. Meteorological data included average daily minimum and maximum temperatures for winter (December to February) and summer (June to September), along with seasonal precipitation for both seasons. Spatial controls included a four-level categorical variable based on US. Census regions (North-East, South, Midwest, and West). The treatment variable was \PM exposure levels for 1982, measured in $\mu g/m^3$ at the census tract level and aggregated to the county level. Specifically, we observe the mean of $17.03 \mu g/m^3$ and the median of $17.47\mu g/m^3$. Binarization of \PMns---which helps reducing the complexity of the exposure which is continuous in nature---has been previously explored and justified in the literature \citep{lee2021discovering, bargagli2020causal, zorzetto2024confounder}. Figure \ref{fig:maps_pm} reports the observed distribution of the \PM level.

\begin{figure}
\begin{center}
\includegraphics[width=4.5in]{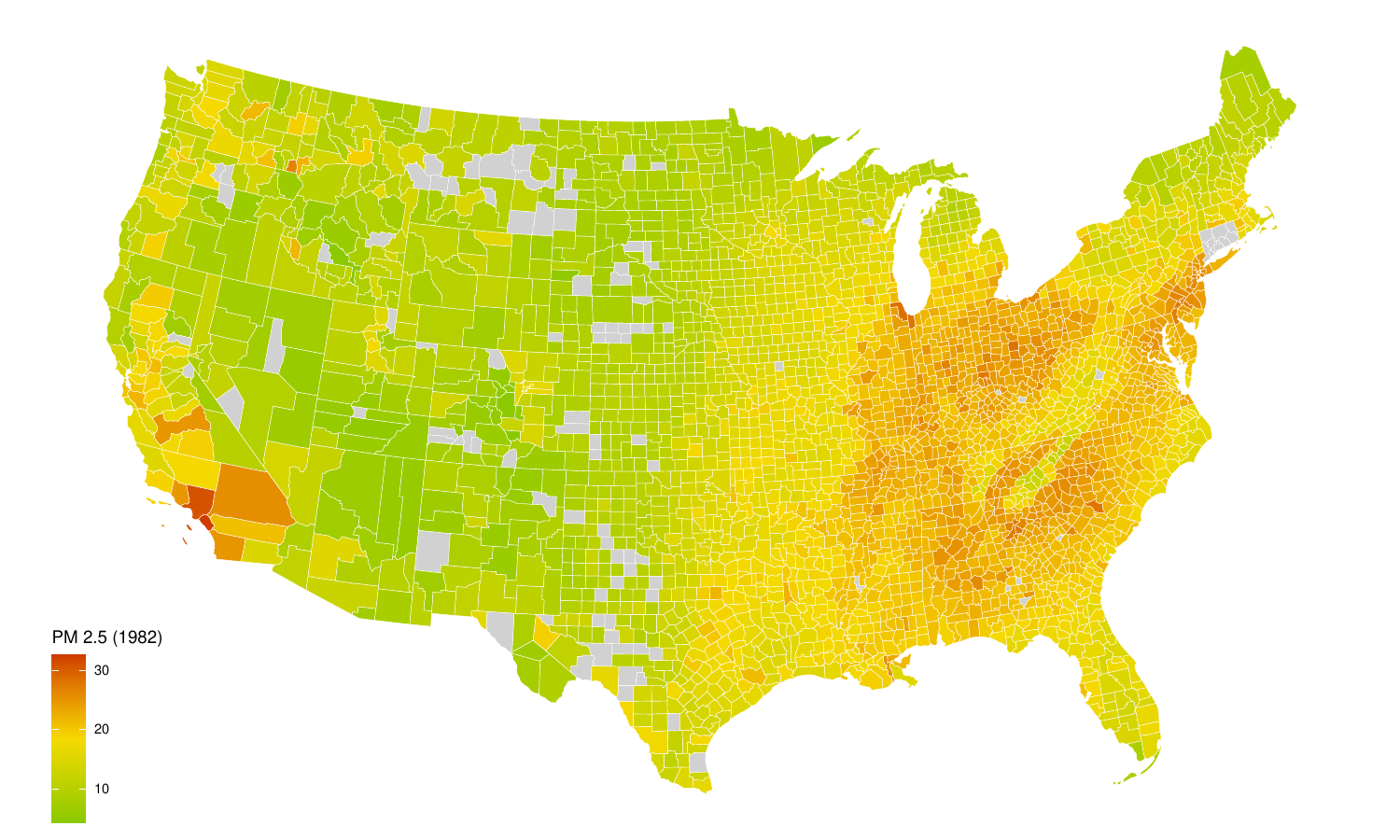}
\end{center}
\caption{Maps of the observed distributions for the levels of \PM during 1982.
}\label{fig:maps_pm}
\end{figure}

\begin{table}
\centering
\small
\begin{tabular}{l|ccc}
\toprule
Variables & Mean & SD & Data source \\
\midrule
Absolute upward mobility (\%)& 43.64 & 6.03 & Opportunity Atlas \\
High school graduation rate (\%)& 79.76 & 5.97 & Opportunity Atlas \\
College graduation rate (\%)& 17.94 & 7.62 & Opportunity Atlas \\
Community college graduation rate (\%)& 28.51 & 9.84 & Opportunity Atlas \\
\PM in 1982 ($\mu g/m^3$) & 17.03 & 5.24 & Opportunity Atlas \\
Share of college-educated in 2000 (\%)& 16.50 & 7.88 & Census \\
Median household income in 1990 (\$) & 24 370.13 & 6 886.06 & Census \\
Population density in 2000 (per sq mile) & 331.10 & 1 049.69 & Census \\
Poverty share in 1990 (\%)& 16.59 & 7.74 & Census \\
Share of black in 2000 (\%)& 9.02 & 14.48 & Census \\
Share of white in 2000 (\%)& 81.86 & 18.37 & Census \\
Share of Hispanic in 2000 (\%)& 6.10 & 11.89 & Census \\
Share of Asian in 2000 (\%)& 0.64 & 1.36 & Census \\
Employment rate in 2000 (\%)& 57.31 & 7.45 & Census \\
Mean winter precipitation (mm/day) & 3.19 & 2.18 & Daymet \\
Minimum winter temperature (°C) & -4.95 & 6.86 & Daymet \\
Mean summer precipitation (mm/day) & 3.23 & 1.46 & Daymet \\
Maximum summer temperature (°C) & 28.95 & 3.32 & Daymet \\
\bottomrule
\end{tabular}
\vspace{0.2cm}
\caption{Descriptive statistics of variables: mean, standard deviation (SD), and source.}
\label{table: descr stats application dataset}
\end{table}

The post-treatment variables include three variables: the community college graduation rate, the high school graduation rate, and the college graduation rate. Chosen for their relevance to the literature on social mobility, these three education levels allow us to study the causal pathway between \PM exposure and social mobility for each of them. The primary outcome is absolute upward mobility \citep[AUM, ][]{chetty2017fading}, which is defined as the mean income percentile in adulthood of individuals born between 1978 and 1983 in families in the $25^{th}$ percentile of the national parent income distribution. Income rank is measured in 2015 (ages 31–37). See Figures C1 and C2 in the Supplementary Materials for the AUM and education attainment distribution, respectively.

A preliminary analysis shows that the community college graduation rate has the strongest correlation with AUM ($66.45\%$) and a significant negative correlation with \PM exposure ($-38.19\%$). High school and college graduation rates also exhibited significant correlations with AUM ($55.55\%$ and $53.47\%$, respectively) and negative correlations with \PM exposure ($-36.32\%$ and $-28.86\%$, respectively). 

\pagebreak
\section{Setup and Causal Estimands}
\label{sec:set_up}

\subsection{Notations and Definitions} We assume to observe $n$ independent and identically distributed units. For each unit $i\in \{1, \dots, n\}$, we let $\X_i \in \mathcal{X} \subseteq \R^q$ be the vector of observed covariates, $T_i \in \{0,1\}$ be the observed binary treatment variable, let $P_i \in \R$ be a post-treatment variable, and $Y_i \in \R$ be a primary outcome of interest. We follow the usual standards of denoting random variables by  capital Roman letters, and their realizations by lower case, while bold letters denote vectors.  

In our application, the units are counties in the continental United States, the treatment variable is the \PM exposure dichotomized with respect to the median of observed values as the threshold, where $t=1$ indicates a high level of \PM exposure and $t=0$ is the low level. The post-treatment variable is a continuous variable that indicates the graduation rate. The primary outcome is social mobility, and the confounders are socioeconomic and meteorological information.

Following Rubin's Causal Model \citep{rubin1974, dominici2020controlled} and invoking the stable unit treatment value assumption (Assumption \ref{sutva}: SUTVA), we postulate the existence of two potential outcomes for the primary outcome, $\{Y_i(0), Y_i(1)\} \in \R^2$, for each $i \in\{ 1, \dots, n\}$, representing the collection of the primary outcomes when the unit $i$ is assigned to the control group---i.e., when $t=0$---or the treatment group---i.e., when $t=1$---, respectively. 

Similarly, following the principal stratification framework \citep{frangakis2002principal}, we also assume the existence of two potential outcomes for the post-treatment variable, $\{P_i(0), P_i(1)\} \in \R^2$, which represent a collection of outcomes for the post-treatment variable when the unit $i$ is assigned to the control or treatment group, respectively.

\begin{assumption}[SUTVA]\label{sutva}
For each unit $i\in \{1, \dots, n\}$, the primary outcome and the post-treatment are a function of their observed treatment level only, such as:
 \begin{eqnarray*} 
& Y_i(T_1, T_2, \cdots, T_i, \cdots,  T_n) = Y_i(T_i)   \mbox{ and } Y_i(T_i) = Y_i;\\
& P_i(T_1, T_2, \cdots, T_i, \cdots,  T_n) = P_i(T_i)  \mbox{ and } P_i(T_i) = P_i.
\end{eqnarray*} 
\end{assumption}
Specifically, SUTVA is a combination of two assumptions: no interference between units---i.e., the potential values of the primary outcome and post-treatment variable of the unit $i$ do not depend on the treatment applied to other units---and consistency---i.e., no different versions of the treatment levels assigned to each unit \citep{rubin1986comment}.
In our applied context, this means that we assume that each county is affected only by the level of \PM in that area and not by the level of \PM in the other counties, since the level of \PM, used in our application, already accounts for geographical confounding.

\subsection{Causal Estimands} 
\label{subsec:causal_est}

Following the contribution of \citet{zigler2012estimating}, we assume that the unit $i$ belongs to the \textit{ associative positive stratum} if $P_i(1)-P_i(0) \geq \xi$, to the \textit{associative negative stratum} if $P_i(1)-P_i(0) \leq -\xi$, or to the \textit{dissociative stratum} otherwise, where $\xi$ is a positive value close to zero. 

The principal causal estimands of interest are respectively the Expected Associated Effect for the positive stratum (EAE$_+$), the negative stratum (EAE$_-$), and the Expected Dissociative Effect (EDE):
\begin{align}
    EAE_+ &=\mathbb{E}\big[Y_i(1)-Y_i(0)\:\:\bigl\rvert\:\: P_i(1)-P_i(0) \geq \xi],\notag \\
    EAE_- &=\mathbb{E}\big[Y_i(1)-Y_i(0)\:\:\bigl\rvert\:\: P_i(1)-P_i(0) \leq -\xi],\label{eq:causal effects} \\
    EDE & =\mathbb{E}\big[Y_i(1)-Y_i(0)\:\:\bigl\rvert\:\: P_i(1)-P_i(0) < |\xi|].
    \notag
\end{align}
%The underlying motivation of causal estimands definition is their relevance and interpretability in the real-data application. 
In our application, the EDE represents the causal effect of exposure to \PM on social mobility in counties where exposure to \PM does not affect educational attainment. In contrast, EAE$_-$ (EAE$_+$) represents the causal effect of exposure to \PM on social mobility given that the counties where exposure to \PM decreases (increases) educational attainment. These estimands allow us to quantify and understand the causal pathways through which \PM exposure affects social mobility, through strata defined based on educational attainment.

In order to identify these causal estimands, we need to introduce the following assumption. 
\begin{assumption}[Strongly Ignorable Treatment Assignment]\label{assumpt2}
For each unit $i \in \{1, \dots, n\}$,
    \begin{gather*}
     \{Y_i(1), Y_i(0), P_i(0), P_i(1) \} \indep  T_i \mid \X_i,\\
      0 < Pr\left(T_i=1 \mid \X_i = \x\right) < 1 \quad \forall \: \x \in \mathcal{X}.
\end{gather*}
\end{assumption}
The strongly ignorable treatment assignment states that the potential outcome for the primary outcome and the post-treatment variable are independent of the treatment conditional on the set of covariates and all units have a positive chance of receiving the treatment. In our application, this means that the potential values of social mobility and educational attainment under the two levels of \PM exposure are independent of the (underlying) mechanism that controls the observed level of \PM in each county, conditional on confounders. Moreover, the values of the confounders for any county cannot preclude the possibility of observing one of the two levels of \PM exposure.

Leveraging the Assumptions \ref{sutva} and \ref{assumpt2}, we can rewrite the principal causal estimand defined in \eqref{eq:causal effects} as the following statistical estimand:
\begin{align}
    {EAE}_+ &= \int_{\x} \Pr(\X_i=\x \mid P_i(1)-P_i(0) \geq \xi)\big(\mathbb{E}\big[Y_i\mid P_i(1)-P_i(0) \geq \xi, T_i=1, \X_i=\x] - \notag \\
    & \quad\quad \mathbb{E}\big[Y_i\mid P_i(1)-P_i(0) \geq \xi, T_i=0, \X_i=\x]\big) d\x; \notag \\
    {EAE}_- &= \int_{\x} \Pr(\X_i=\x \mid P_i(1)-P_i(0) \leq -\xi)\big(\mathbb{E}\big[Y_i\mid P_i(1)-P_i(0) \leq -\xi, T_i=1, \X_i=\x] - \notag \\
    & \quad\quad \mathbb{E}\big[Y_i\mid P_i(1)-P_i(0) \leq -\xi, T_i=0, \X_i=\x]\big) d\x; \notag \\
    {EDE} &= \int_{\x} \Pr(\X_i=\x \mid P_i(1)-P_i(0) < \mid\xi\mid)\big(\mathbb{E}\big[Y_i\mid P_i(1)-P_i(0) < \mid\xi\mid, T_i=1, \X_i=\x] - \notag \\
    & \quad\quad \mathbb{E}\big[Y_i\mid P_i(1)-P_i(0) < \mid\xi\mid, T_i=0, \X_i=\x]\big) d\x.
    \label{eq:stat_estimand}
\end{align}
Let us indicate with $f(P_i(1),P_i(0);\xi)$ the general definition of the three strata---e.g. for the positive associative stratum $f(P_i(1),P_i(0);\xi)= P_i(1)-P_i(0) \geq \xi$. 
Then, the inner expectations in each statistical estimand $\E\big[Y_i \mid T_i=t,  \X_i=\x, f(P_i(1),P_i(0);\xi) \big]$ can be estimated with the outcome model $\{Y_i \mid T_i, \X_i, P_i(1),P_i(0)\}$ (which is defined in \eqref{eq:mixture-model} in Section \ref{sec:model}). Similarly, the probabilities in \eqref{eq:stat_estimand} can be rewritten as follows:
\begin{equation*}
    \Pr(\X_i=\x \mid f(P_i(1), P_i(0);\xi))= \frac{\Pr(f(P_i(1), P_i(0);\xi) \mid\X_i=\x)\Pr(\X_i=\x)}{\int_{\x}\Pr(f(P_i(1), P_i(0);\xi) \mid\X_i=\x) d\x}
\end{equation*}
where $\Pr(f(P_i(1), P_i(0);\xi) \mid\X_i=\x)$ is estimated via the potential post-treatment variables model (defined in \eqref{eq:P_model}), while the $\Pr(\X_i)$ is observed.

\section{Bayesian Semi-parametric Approach}
\label{sec:model}

\subsection{Model Formulation}
\label{subsec:model_def}

Following the Bayesian paradigm and assuming SUTVA, the joint distribution of confounders $\X$, treatment $T$, potential outcome for post-treatment variable $(P(0),P(1))$, and potential outcome for the primary outcome $(Y(0),Y(1))$ can be rewritten as follows:
$$\Pr(\X,T, P(0), P(1),Y(0), Y(1)
)=\int_\Theta \prod_{i=1}^n \Pr(X_i,T_i, P_i(0), P_i(1),Y_i(0), Y_i(1)\mid \theta) p(\theta) d\theta, $$ 
where the inner probability distributions  
can be factorized into:
\begin{align}
    \Pr( T_i \mid Y_i(0), Y_i(1), P_i(0), P_i(1), \X_i, \theta_t) & \times \Pr(Y_i(0), Y_i(1) \mid P_i(0), P_i(1), \X_i, \theta_y) \notag \\
     \times \Pr(P_i(0), P_i(1)  \mid \X_i,      \theta_p)  & \times \Pr( \X_i \mid \theta_x);
    \label{eq:decom_prob}
\end{align}
where $\theta = \{\theta_t,\theta_y,\theta_p,\theta_x\}$ and $p(\theta)$ is the prior distribution for all the parameters $\theta$, which take values in the parameter space $\Theta$.
For a compact notation, we use the symbol $\Pr(\cdot)$ to denote the probability law.

The strong ignorability assumptions, defined in Section \ref{sec:set_up}, allow us to simplify the conditional probability for the treatment variable as dependent only on the confounders,
that is, $\Pr(T_i \mid Y_i(0), Y_i(1), P_i(0), P_i(1), \X_i, \theta_t) = \Pr(T_i \mid \X_i, \theta_t)$. 

Following \cite{schwartz2011bayesian}, the treatment and covariate distributions are directly observed, thus they are not needed 
to be modeled; while we need to model the conditional distribution of the potential outcome of the post-treatment variable, given the confounders and treatment, and the conditional distribution of the potential outcome for the primary outcome, given the potential outcome for post-treatment variable, the confounders and the treatment.

We use a nonparametric mixture to model the conditional distribution of the primary outcome. Specifically, we leverage the dependent Dirichlet process \citep[DDP ---][]{mac2000dependent, barrientos2012support, quintana2020dependent}, assuming that,
given the treatment level $t\in \{0,1\}$, %for each unit $i=1,\dots n$ 
\begin{gather}  \label{eq:mixture-model}
Y_{i} \mid \x_i,t,p_{i}(0),p_{i}(1), H_{\x_i,p_{i}(0),p_{i}(1)}^{(t)} \overset{indep}{\sim} 
\int_{\Theta_y} {\mathcal Q} \big(\cdot \mid \x_i,p_{i}(0),p_{i}(1), \theta_y \big) d H_{\x_i,p_{i}(0),p_{i}(1)}^{(t)}(\theta_y), 
\end{gather}
for units $i=1, \ldots, n$, where $\mathcal Q$ is a continuous kernel density for every $\theta_y \in \Theta_y$ that depends on both the confounders $\x$ and on the potential outcome for post-treatment variables $\big(p(0),p(1)\big)$; and the collection of random distributions $\{ H_{\x_i,p_i}^{(t)}\}$ --- for varying confounders levels $\x$ and $p(0), p(1)$ and fixed $t$--- is given a DDP prior law. While, in model (\ref{eq:mixture-model}), the $Y_i$ are conditionally independent, probabilistic dependence across them, thus borrowing strength across counties, is implied by the DDP prior law on the $\{ H_{\x_i,p_i}^{(t)}\}$. 
Specifically, the DDP prior we propose leverages on the stick-breaking representation of the DP \citep{sethuraman1994constructive}, letting
\begin{equation*}
H_{\x_i,p_i(0),p_i(1)}^{(t)}(\cdot)  = \sum_{m=1}^M \lambda^{(t)}_m(\x_i) \delta_{{\mathbb{\theta}^{(t)}_m(\x_i,p_i(0),p_i(1))}}(\cdot),
\end{equation*}
where the infinite sequences of random weights $\{\lambda^{(t)}_m(\cdot)\}_{m\geq1}$ and of random atoms $\{\mathbb{\theta}^{(t)}_m (\cdot)\}_{m\geq1}$ are stochastic processes; and in our specification, both stochastic processes depend on the confounders $\x$, and the random atoms also depend on the post-treatment variables. 

This {specification gives a flexible nonparametric mixture model for each $i$ such that 
$$  
Y_{i} \mid \x_i,t, p_i(0), p_i(1), \{\lambda_m^{(t)}, \theta_m^{(t)}\}  
\overset{indep}{\sim}
 \sum_{m=1}^\infty  \lambda_m^{(t)}(\x_i) \; 
  {\mathcal Q} \big(y ;\x_i, p_i(1),p_i(0), \mathbb{\theta}_m^{(t)}(\x_i,p_i(1),p_i(0)) \big).
$$ 
The model so defined for the primary outcome distribution allows the corresponding potential outcomes to depend on both confounders and post-treatment variable, which is essential for a flexible model to correctly estimate the principal causal effects. Moreover, importantly, the specific choice of including the confounders in the infinite sequences of the weights allows us to characterize the heterogeneity in the potential outcomes for primary outcome and improve the imputation of the missing data, as shown in \citet{zorzetto2024confounder} in the case of estimation of the heterogeneous treatment effect. See \citet{wade2025bayesian} for a broad discussion on incorporating dependence either through the random weights or the random atoms of the DDP in more general settings.

Following the Probit stick-breaking process (PSBP) introduced by \citet{rodriguez2011nonparametric} and adapted for the causal inference setting, the random weights $\{\lambda^{(t)}_m(\x_i)\}_{m\geq1}$, for each unit $i$, can be defined as follows:
\begin{gather}
    \lambda^{(t)}_m(\x_i)  = \Phi(\gamma^{(t)}_m(\x_{i}))\prod_{a<m}\{1-\Phi(\gamma^{(t)}_a(\x_{i}))\}, \notag \\
    \gamma^{(t)}_m(x_{i})\mid \epsilon^{(t)}, \sigma_{\gamma}^2 \overset{indep}{\sim}  N([1, x_i]' \epsilon_m,\sigma_{\gamma}^{2}), 
\label{eq:model_lambda}
\end{gather}
where $\Phi(\cdot)$ denotes the cumulative distribution function of a standard Normal distribution and the $\gamma^{(t)}_m(\x)$ have independent Gaussian distributions with a linear combination of the confounders $\x$ as the mean, such that $\epsilon^{(t)}_m$ are the regression parameters, including the intercept, for each cluster $m$ and treatment level $t$.

The weights $\{\lambda^{(t)}_m(\x)\}_{m\geq1}$ implicitly describe the probability of belonging to each cluster, under treatment $t$ and given confounders $\x$. Therefore, we can introduce a latent categorical variable $V_i^{(t)}$, for each unit $i \in\{1, \dots, n\}$ and each treatment level $t\in \{0,1\}$, such that 
\begin{gather*}
    Pr(V_{i}^{(t)}=m \mid \x_{i}) =\lambda^{(t)}_m(\x_i),\quad  m= 1, 2,\dots \notag
\end{gather*}

Given a specific cluster allocation $V_{i}^{(0)}=m_0$ and $V_{i}^{(1)}=m_1$ respectively under the two treatment levels, and assuming that the kernel $\mathcal{Q}(\cdot)$ is a Gaussian distribution, we can rewrite the model \eqref{eq:mixture-model} for the two treatment levels as follows:
\begin{gather}
    Y_{i}(0) \mid \x_{i},p_i(0),p_i(1),\theta_y^{(0)},
    (V_{i}^{(0)}=m_0) \sim {\mathcal N}\bigg([1, \x_i,p_i(0)]'\boldsymbol{\eta}^{(0)}_{m_0},{\sigma_{y,m_0}^{(0)}}^2\bigg),\notag \\
     Y_{i}(1) \mid \x_{i},p_i(0),p_i(1),\theta_y^{(1)}, (V_{i}^{(1)}=m_1) \sim {\mathcal N}\bigg([1, \x_i, p_i(1),p_i(0)]'\boldsymbol{\eta}^{(1)}_{m_1},{\sigma_{y,m_1}^{(1)}}^2\bigg),
   \label{eq:Ynormal_cluster}
\end{gather}
where $(\eta^{(t)}_m)_{m\geq 1}$ are regression parameters and $(\sigma_{y,m}^{(t)})_{m\geq 1}$ are scale parameters for $t=0, 1$ and we let $\theta_y^{(t)} =(\theta_{y,m}^{(t)})_{ m\geq 1} =\big((\boldsymbol{\eta}^{(t)}_m, {\sigma_{y,m}^{(t)}}^2)\big)_{ m\geq 1}$. We assume the following prior distributions for each treatment $t$ and cluster $m$
\begin{equation*}
\eta_m^{(t)}\sim \Norm(\mu_\eta, \sigma^2_\eta)\quad\mbox{and}\quad {\sigma_{y,m}^{(t)}}^2 \sim \mbox{InvGamma}(\gamma_1,\gamma_2).
\end{equation*}
Our model corresponds to what is often referred to as \textit{T-learner} in the causal inference literature \citep{li2022bayesian}.

Moreover, for the potential outcome for post-treatment variables distribution, we assume a linear regression with Gaussian errors, such that the post-treatment variables are independent given the treatment $t$ and dependent to the confounders $\boldsymbol{x}$
\begin{equation}
    P_i(t)\mid \boldsymbol{x}_i, \theta_p \sim \mathcal{N}([1,\boldsymbol{x}_i]'\boldsymbol{\beta}^{(t)}, {\sigma_p^{(t)}}^2); \label{eq:P_model}
\end{equation}
where $ \theta_p=\{\boldsymbol{\beta}^{(t)},{\sigma_p^{(t)}}^2\}_{t=0,1}$. We assume a conjugate prior for the linear regression parameters $\boldsymbol{\beta}^{(t)}\sim \Norm_{q+1}(\mu_\beta, \sigma^2_\beta \mathbf{I}_{q+1})$ and, for the variance, ${\sigma_p^{(t)}}^2 \sim \mbox{InvGamma}(\gamma_3,\gamma_4)$.

\subsection{Model's Properties}
Although the choice of a parametric distribution for the potential outcomes of the post-treatment variable in \eqref{eq:P_model} may appear restrictive compared to the more flexible specifications used in principal stratification approaches with Bayesian nonparametric methods \citep[e.g.,][]{schwartz2011bayesian, zorzetto2024bayesian, antonelli2023principal}, our focus lies on the predictive distribution, meant as the conditional distribution of the missing post-treatment variable given the observable variables, which, as we are going to show, is a mixture distribution capable of capturing complex and flexible patterns in the post-treatment variable.

As outlined in Section \ref{sec:set_up}, the potential outcome framework inherently involves a missing data problem: only one potential outcome is observed for the post-treatment variable for each unit, the other one is counterfactual. Consequently, in computations, and in particular in the Gibbs sampling procedure---detailed in the Supplementary Material---we include an imputation step for the unobserved post-treatment variable, drawing from its predictive distribution.

From Bayes theorem, the general expression (with some abuse of notation) of the predictive distribution for the post-treatment variable is
\begin{align*}
    \Pr(P(0), P(1)\mid \boldsymbol{x},Y(0), Y(1)) = \int_\Theta &\frac{\Pr(Y(0), Y(1)\mid \boldsymbol{x},P(0), P(1), \theta)\Pr(P(0), P(1)\mid \boldsymbol{x}, \theta)}{\Pr(Y(0), Y(1)\mid \boldsymbol{x},\theta)}\\
    &\times\Pr(\theta\mid Y(1),Y(0),P(1),P(0))d\theta;
\end{align*}
where $\Pr(Y(0), Y(1)\mid \boldsymbol{x},P(0), P(1), \theta)$ denotes the distribution of the primary outcome, in our case defined in Eq \eqref{eq:mixture-model}; $\Pr(P(0), P(1)\mid \boldsymbol{x}, \theta)$ refers to the distribution of the post-treatment variable, as defined in equation \eqref{eq:P_model}; and the denominator can be regarded as a constant, since it does not depend on $P(0)$ or $P(1)$.

Therefore, we can write the predictive distribution conditional on the parameters according to our model assumptions and given the observed variable, for each unit $i$ and its missing treatment $1-t$, as follows 
\begin{align}
& \Pr\left(P_i(1-t)\mid T_i=t,\boldsymbol{x}_i,Y(0), Y(1),\boldsymbol{\beta}^{(1-t)},{\sigma_p^{(1-t)}}^2,\boldsymbol{\eta}^{(t)},\boldsymbol{\sigma}_y^{(t)}\right) \label{eq:post_pred}\\
&\quad\quad \propto \mathcal{N}\left(P_i(1-t);\boldsymbol{\beta}^{(1-t)}\boldsymbol{x}_i,{\sigma_p^{(1-t)}}^2\right)\sum_{m=1}^\infty\lambda^{(t)}_m(\boldsymbol{x}_i)\mathcal{N}\left(Y_i(t);\boldsymbol{\eta}^{(t)}_m\boldsymbol{x}_i, \sigma_{y,m}^2\right)\notag \\
    %P_i(t)|X_i,t,Y(1-t),\beta^{(t)},{\sigma^{(t)}}^2,\eta^{(t)},\sigma_{y^{(t)}} \sim \mathcal{N}(\beta^{(t)}X_i^p,{\sigma^{(t)}}^2)\sum_{m=1}^M\lambda^{(1-t,m)}(X_i)\mathcal{N}(\eta^{(1-t,m)}X^{(y_1-t)}, {\sigma_{y_1-t}^{(m)}}^2) \\
     &\quad\quad= \sum_{m=1}^\infty\lambda^{(t)}_m(\boldsymbol{x}_i)\left[\mathcal{N}\left(P_i(1-t);\boldsymbol{\beta}^{(1-t)}\boldsymbol{x}_i,{\sigma_p^{(1-t)}}^2\right)\mathcal{N}\left(Y_i(t);\boldsymbol{\eta}^{(t)}_m\boldsymbol{x}_i, \sigma_{y,m}^2\right)\right] \notag\\ 
     %&\propto \sum_{m=1}^M \lambda^{(1-t)}_m(x_i)Pr(P_i(t)|V_i(1-t)=m\ldots) 
     &\quad\quad= \sum_{m=1}^\infty \lambda^{(t)}_m(\boldsymbol{x}_i)\Pr\left(P_i(1-t)\mid T_i=t,V_i(t)=m, \boldsymbol{x}_i,Y_i(t),\boldsymbol{\beta}^{(1-t)},{\sigma_p^{(1-t)}}^2,\boldsymbol{\eta}^{(t)},\boldsymbol{\sigma}_{y}\right).
     \notag
\end{align}
where $\Norm(Z;\mu, \sigma^2)$ denotes the Gaussian density distribution with parameters $\mu$ and $\sigma^2$ for the random variable $Z$.
The term $\Pr(P_i(1-t)\mid T_i=t,V_i(t)=m, \boldsymbol{x}_i,Y_i(t), \theta)$ denotes the probability distribution of the post-treatment variable under the unobserved treatment level $t$, given the cluster allocation $m$ of the observed primary outcome. By leveraging properties of the Gaussian distribution, this quantity can be reformulated as follows:

\begin{gather*}
    P_i(1-t)\mid T_i=t,V_i(t)=m, \boldsymbol{x}_i,Y_i(t), \theta \sim \mathcal{N}\left( \mu_{i,m}, \tau_{i,m}^{-1} \right),\\
    \mu_{i,m} = \frac{\boldsymbol{\beta}^{(1-t)}\boldsymbol{x}_i\sigma_{y}^2+{\sigma_p^{(1-t)}}^2\eta_{m,2}^{(t)}\biggl[Y_i(t)-\eta_{m,0}^{(t)}-\eta_{m,1}^{(t)}\boldsymbol{x}_i-\eta_{m,3}^{(t)}P_i(t) \biggl]}{\sigma_{y,m}^2+{\sigma^{(1-t)}_p}^2\eta_{m,2}^{(t)}},\\
    \tau_{i,m}= \frac{1}{{\sigma^{(1-t)}_p}^2} + \frac{\left({\eta_{m,2}^{(t)}}\right)^2}{\sigma_{y,m}^2}.
\end{gather*}

Therefore, conditional on the parameters, the predictive distribution for the counterfactual post-treatment variable in \eqref{eq:post_pred} can be expressed as a mixture of Gaussian distributions, where the regression means depend on the observed values of the post-treatment variable and the primary outcome. This formulation highlights that, despite the parametric assumptions imposed on the likelihood of the post-treatment variable, the resulting predictive distribution remains highly flexible and capable of capturing complex distributional features.
\section{Simulation Study}
\label{sec:simulations}

\subsection{Data Generating Process}

The simulation study is designed to allow us to test our proposed model's performance to impute the missing post-treatment variables and the missing potential values for the primary outcome which is crucial to correctly estimate the causal effects. To evaluate that we estimate the bias for the average treatment effect in the simulated sample for the potential outcome for post-treatment variable and for the primary outcome, respectively $ATE_P = \frac{1}{n} \sum_{i=1}^{n} P_i(1)-P_i(0)$ and $ATE_Y = \frac{1}{n} \sum_{i=1}^{n} Y_i(1)-Y_i(0)$
where $n$ is the sample size of each generated scenario.

We evaluate our proposed model under three data generating processes. These three different scenarios vary in the complexity of the distributions of post-treatment variables and primary outcomes.  Scenario 3 mimics the characteristics of the dataset analyzed in the data application in Section \ref{sec:application}. The results are compared with the semi-parametric model for principal stratification introduced by \citet{schwartz2011bayesian} and the copula model introduced by \citet{lu2023principal}.

For each of the following three Scenarios, we consider 200 replicates. Each replicate has $500$ units in Scenarios 1 and 2, and $300$ units in Scenario 3---similarly to dataset analyzed in Section \ref{sec:application}.

\vspace{0.2cm}
\underline{\it Scenario 1}: We define five confounders $X_{1:5}$---two Bernoulli random variables and three standard Gaussian random variables---and a Bernoulli treatment variable which depends on the three Gaussian random variables, $T \sim \mbox{Be}(f_1(X_{3:5}))$. As in Equation~\eqref{eq:simulation}, the potential outcome for the post-treatment variable is sampled from a Gaussian linear regression with a mean that depends on covariates $X$ and treatment-specific coefficients, while the potential outcomes for the primary outcome are sampled from a mixture of Gaussian distributions.    
The units are divided into three clusters $m \in \{1,2,3\}$ according to the values of the confounders $X_{1:2}$---see further details in the Supplementary Material---, which determine the allocation to the different components in the mixture. The distributions of the potential outcomes for post-treatment variables and primary outcomes are defined as follows:
\begin{gather}
     P_i(0)\sim \Norm\Big(\beta^{(0)} {\bf X},\sigma_{p}^2\Big),\quad 
    P_i(1)\sim \Norm\Big(\beta^{(1)} {\bf X},\sigma_{p}^2\Big), \notag\\
    Y_i(0)\sim \sum_{m=1}^3\mathbb{I}_{(M_i=m)}\Norm\Big(\eta_m^{(0)} g_{0}({\bf X},P(0)),\sigma_{y,m}^{(0)2}\Big), \label{eq:simulation} \\
    Y_i(1)\sim \sum_{m=1}^3\mathbb{I}_{(M_i=m)}\Norm\Big(\eta_m^{(1)} g_{1}({\bf X},P(0),P(1)),\sigma_{y,m}^{(1)2}\Big);
    \notag
\end{gather}
for each unit $i \in \{1, \dots, n\}$, where the function $g_{0}(\cdot)$ and $g_{1}(\cdot)$ are nonlinear functions different for the two treatment levels, and $\mathbb{I}_{(\cdot)}$ is an indicator variable. The variable $M_i$ indicates the cluster allocation for each unit $i \in \{1, \dots, n\}$.

\vspace{0.2cm}
\underline{\it Scenario 2}: We increase the number of confounders to $10$ variables. The treatment level is defined as $T \sim \mbox{Be}(f_2(X_{1,2,6}))$. Cluster allocation depends on the binary confounders $X_{1:2}$ as in Scenario 1, while the potential outcome for post-treatment variable distribution depends on the remaining confounders $X_{3:10}$. By assigning different sets of confounders to the cluster allocation and the post-treatment potential outcomes, we introduce a distinct heterogeneity scheme compared with Scenario 1. The potential outcomes for the primary outcome distribution are defined as in equation~\eqref{eq:simulation}, with all $10$ confounders included in the regression of the means within the cluster-specific Gaussian mixture.

\vspace{0.2cm}
\underline{\it Scenario 3}: This Scenario closely mimics the characteristics of the dataset used in the application Section \ref{sec:application}. We reduce the number of units and increase the number of covariates to 14, including Bernoulli distributed and normally distributed variables with varying variances. The treatment variable, the cluster allocation, and the potential outcome for the post-treatment variable distribution are defined similarly to Scenario 2 but incorporate a larger number of covariates.

\vspace{0.2cm}
The Gibbs sampler, implemented in \texttt{R} and available in \texttt{GitHub} at \href{https://github.com/dafzorzetto/StrataBayes_SocialMobility}{\texttt{dafzorzetto/StrataBayes\_SocialMobility}}, allows the users to define distinct sets of covariates for the regression for the post-treatment variable, the regression for the primary outcome, and the regression characterizing the weights---i.e., the covariates that describe the cluster allocation.

\subsection{Results}
Table \ref{table: simulation_bias} reports the comparison of the bias for $ATE_P$ and $ATE_Y$ between our proposed model (denoted Y\_BNP), the semi-parametric model by \citet{schwartz2011bayesian} (indicated with SLM), and the copula model by \citet{lu2023principal} (indicated with LJD). The latest model allows us to estimate only the $ATE_Y$, because the algorithm does not impute the missing post-treatment variable. The median and the interquartile range (IQR) are estimated in the $200$ replicates for each scenario. %Additional visualizations are provided in the supplementary materials.

Table \ref{table: simulation_bias} shows that our proposed Y\_BNP model consistently outperforms the SLM and the LJD model in terms of both bias and variability across all scenarios. Specifically, Y\_BNP produces median estimates closer to zero and smaller IQRs for both causal quantities,  $ATE_P$ and $ATE_Y$, in each of the three scenarios. 

The results for $ATE_P$ show that the Y\_BNP model effectively captures the heterogeneity in the post-treatment variable distribution and accurately imputes missing data. Across all scenarios, the median bias ranges from -0.0129 to -0.0182, with narrow IQRs that include the zero. 
This indicates robust performance of the Y\_BNP model on the post-treatment variable, avoiding substantial over- or underestimation, even though the model assumes a linear regression structure. As previously discussed, the predictive posterior distribution of the potential outcome for post-treatment variable is not a single Gaussian distribution with linear regression but a mixture of them, allowing for greater flexibility. In contrast, the SLM model exhibits a significantly higher median bias, particularly in Scenarios 2 (0.4037) and 3 (0.4340), with IQRs nearly three times higher than those of the Y\_BNP model.

%indicating a wide spread in estimates and potential instability.
%These results suggest that the SLM model may be more prone to inaccuracies in these Scenarios.

The difference in performance is further underscored by the results for the $ATE_Y$.
The Y\_BNP model demonstrates greater accuracy compare with both SLM and LJD models. Across the three scenarios, our proposed model has small median bias values ranging from -0.0333 to -0.0896, and the IQRs taking values between 0.0818 and 0.1790. In contrast, the SLM model exhibits a significantly larger bias and interquartile range, in particular for Scenario 1 with median 160.8045 and IQR 10,121.2774. In Scenarios 2 and 3, both the compared model---SLM and LJD---have larger bias with median ten times the one estimated by the Y\_BNP model and the IQR is more than double of the one estimated by the Y\_BNP model. The particularly high values observed in Scenario 1 highlight SLM model's difficulty in capturing complex heterogeneity in the data and adequately controlling the propagation of variability from the post-treatment variable to the primary outcome. While, the SLM model incorporates flexibility in modeling  the post-treatment variable, this alone is insufficient due to its reliance on a linear model for the primary outcome distribution. The LJD model marks an improvement with respect to the SLM but still underperforms with respect to the Y\_BNP.  Its bias values (–0.5642 to –0.2682) and IQRs (0.3005 to 0.6148) remain significantly higher, typically an order of magnitude greater in bias and at least twice as wide in variability compared to the Y\_BNP. 

%Accurate and stable estimation of treatment effects is particularly important in the context of principal stratification analysis, where both the missing post-treatment variable and the potential outcome for primary outcomes have to be imputed. Leveraging a Bayesian nonparametric prior to the primary outcome provides a more robust and reliable framework for capturing heterogeneity in the data and controlling the propagation of variability that often arises in this type of scenario.

\begin{table}
\centering
\begin{tabular}{llrccc}
 &&& Scenario 1 & Scenario 2 & Scenario 3 \\
  \toprule
  \multicolumn{2}{l}{Bias ATE\_P (Post-treatment variable)} &&&& \\
  \midrule
Y$\_$BNP & Median && -0.0129  & -0.0129  & -0.0182  \\ 
 & IQR && 0.1020  & 0.0417  & 0.0604  \\ 
SLM & Median && -0.1419  & 0.4037  & 0.4340 \\ 
 & IQR &&  0.3134 &  0.1344  & 0.1690 \\ 
 LJD &  Median && -- & -- & -- \\ 
 & IQR && -- & -- & --  \\ 
 \toprule
\multicolumn{2}{l}{Bias ATE\_Y (primary outcome)} &&&& \\
  \midrule
Y$\_$BNP & Median && -0.0333  & -0.0377  & -0.0896  \\ 
 & IQR && 0.1400  & 0.0818  & 0.1790  \\  
SLM & Median &&  160.8045 & -0.6845 & -1.0133 \\ 
 & IQR &&  10121.2774  & 0.5069  & 0.7639 \\
 LJD &  Median &&  -0.5642 & -0.2773 & -0.2682 \\ 
 & IQR && 0.6148 & 0.3005 & 0.3603  \\ 
\bottomrule
\end{tabular}
\vspace{0.2cm}
\caption{Comparison of the median and interquartile range (IQR) of the bias for the Sample Average Treatment Effect for post-treatment variable and primary outcome---i.e., $ATE_P$ and $ATE_Y$---between our proposed model (Y$\_$BNP), the \citet{schwartz2011bayesian}'s model (SLM), and the \citet{lu2023principal}'s model (LJD).}
\label{table: simulation_bias}
\end{table}

\section{Empirical Application}
\label{sec:application}

In this section, our goal is to address  the research question: \textit{What is the causal effect of \PM exposure on social mobility within principal strata defined by potential educational attainment under different \PM exposure scenarios?}

More specifically, we identify three main categories to define educational attainments: \textit{high school, community college,} and \textit{college}. In the US context, the difference between community college and college lies in both academic offerings and accessibility of the two, where the former generally focuses on imparting practical skills that prepare students for the workforce or for further education---usually in the form of two-year degrees---and the latter focuses on academic teaching and research in the form of four-year bachelor’s degrees. 

It is important to note that, contrarily to colleges, community colleges are more accessible to a larger portion of the population due to their lower tuition fees. Therefore, we replicate the analysis where the post-treatment variable varies, assuming each of the three education levels, while keeping constant the \PM exposure as treatment, the social mobility as primary outcome, and socio-economic characteristics as confounders.

Subsection \ref{subsec:data} describes the data used for this analysis, obtained merging and harmonizing different data sources. The following subsections report the empirical results of the principal stratification analysis for the three education levels estimated by our proposed model and the characterization of the identified principal strata.

\subsection{Principal Causal Effects}
\label{subsec:results_1}
%The post-treatment variables are normally distributed as tested visually by looking at their histograms and their Q-Q plots. This is important to justify the Gaussian distribution assumed for the post-treatment variable model.

We applied our proposed model in Section \ref{subsec:model_def} to the constructed dataset. Specifically, we replicate the analysis for the three education levels: high school, community college, and college. The three analyses share the same set of confounders---i.e., socio-economic characteristics, meteorological information as well as the spatial confounder---; the treatment, that is defined as exposure to high or lower level of \PMns---the lower (high) level when \PM is below (above) the threshold of $17.5\mu g/m^3$---; and the primary outcome represented by social mobility which is measured by AUM. 

We assume that the threshold $\epsilon$  for the principal strata, defined in \eqref{eq:causal effects}, is equal to $0.01$ for the three education rates. This choice is justified by a change in the level of education that is deemed notable considering the range of the estimated difference between the education rates under treatment and under control. Table \ref{table: balance education, strata} shows that the three principal strata---identified for each education variable after estimating our proposed model---exhibit similar proportions of units in the three post-treatment variables. 

More than $50\%$ of the counties fall into the associative negative stratum, indicating that for these cases, the \PM exposure has a negative effect on the education level. In this stratum, the educational attainment rate in counties exposed to a higher level of \PM is significantly lower than it would have been exposed to a lower \PM level. This observation aligns with findings previously reported in the literature, supporting its plausibility \citep[see e.g.,][]{brochu2011,Bell2012,Binelli2015,Hajat2015,Grunewald2017,Yang2018}. Among the counties that were considered, between $10.4\%$ and $14.8\%$, across the three education levels, are included in the dissociative stratum---i.e., the counties where we do not found evidence of an effect of \PM on the education level. While, in all three education levels, approximately $30\%$ of units belong to the associative positive stratum, the stratum that includes counties that experience increased education rates when exposed to higher levels of \PMns.

\begin{table}
\centering
\begin{tabular}{l|ccc}
\toprule
 & Associative negative & Dissociative & Associative positive \\ 
\midrule
Community college & 56.7\% & 10.5\% & 32.8\% \\ 
High school & 56.3\% & 14.8\% & 28.9\% \\ 
College & 53.1\% & 13.6\% & 33.3\% \\ 
\bottomrule
\end{tabular}
\vspace{0.2cm}
\caption{Percentage of units belonging to each stratum: associative negative stratum, dissociative stratum, and associative positive stratum, across three different educational post-treatment variables: community college, high school, and college.}
\label{table: balance education, strata}
\end{table}

Figure \ref{fig:application results} reports the posterior distribution of the principal causal effects of \PM exposure on social mobility (AUM) using as post-treatment variable the education level, respectively high school, community college, and college.

The posterior distributions of the $EAE_-$ have mean -3.9\%, -5.0\%, and -4.7\% for the analysis with high school, community college, and college as post-treatment variables, respectively. These results indicate that in counties where higher \PM exposure significantly reduces educational attainment, social mobility also decreases by approximately 5\%. The 95\% credible intervals for the causal effect within the negatively associated strata are $(-5.6\%,-2.1\%)$ when using high school attainment as the post-treatment variable, $(-6.8\%,-3.1\%)$ with community college, and $(-6.5\%,-2.8\%)$ in the case of using college.

The posterior distributions of the $EDE$ further support the negative effect of the \PM exposure on social mobility. Specifically, the posterior means for high school, community college, and college as post-treatment variables are -2.3\%, -1.5\%, -2.0\%, respectively. The corresponding 95\% credible intervals are $(-4.4\%,-0.3\%)$, $(-3.6\%,0.7\%)$, and $(-4.0\%,0\%)$. Notably, the credible interval includes zero only in the case of community college. These findings indicate that even in counties where the education level is not affected by the \PM exposure---i.e., within the dissociative stratum---the social mobility is reduced when exposed to higher levels of air pollution.

Lastly, we estimated posterior distributions for $EAE_+$, which results to be centered around zero. Specifically, the posterior means are -0.9\%, 0.3\%, and 0\%, for high school, community college, and college, respectively, with 95\% credible intervals that include the zero across all three education variables. These results suggest that for the counties where higher \PM exposure is associated with increased educational attainment, social mobility does not appear to be significantly affected by air pollution exposure.

\begin{figure}
\begin{center}
\includegraphics[width=5in]{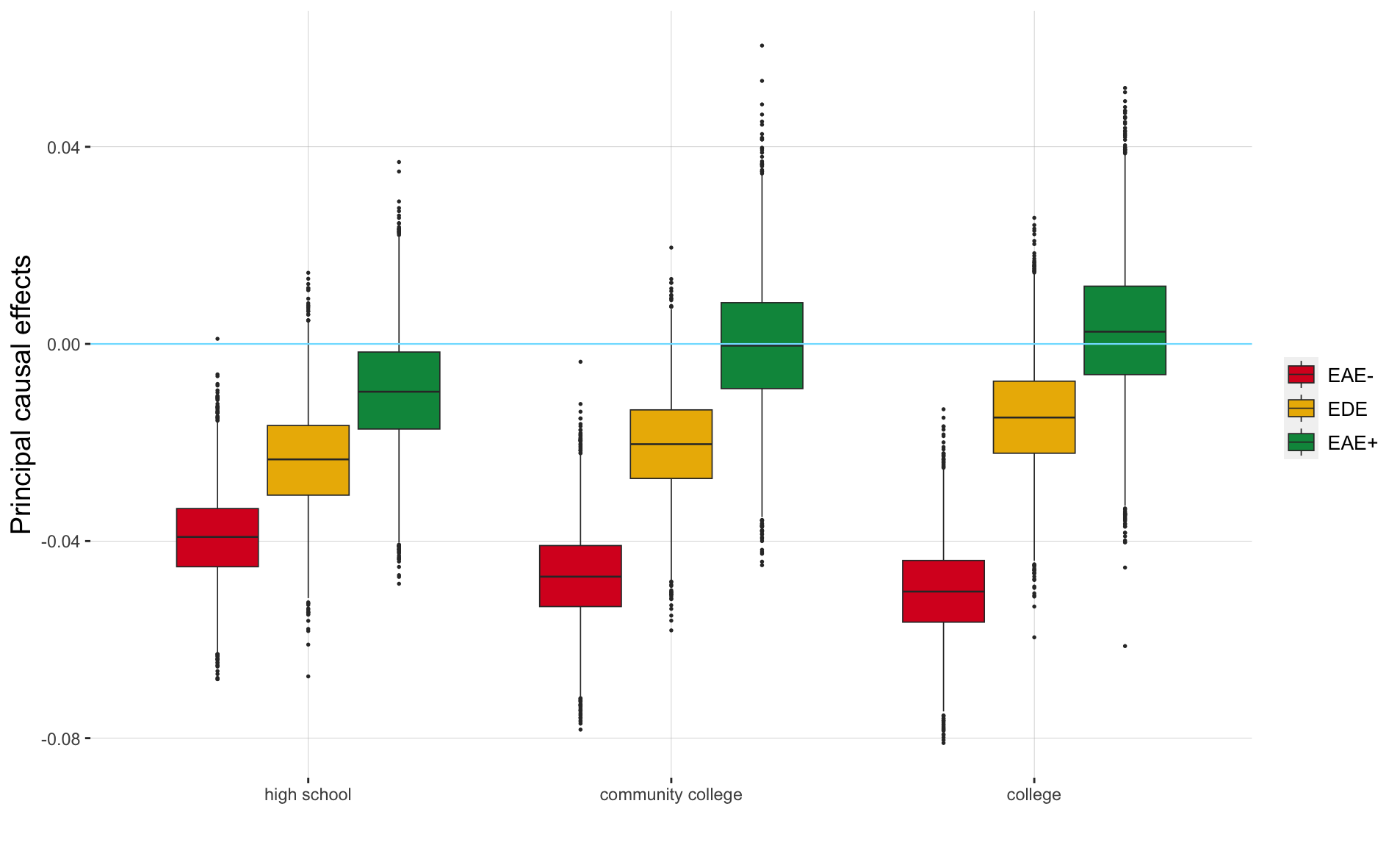}
\end{center}
\caption{Posterior distribution of the principal causal effects of \PM exposure on social mobility (AUM) using as post-treatment variable the education level, respectively high school, community college, and college. In red the Expected Associated Effect for the positive stratum (EAE$_+$), in yellow Expected Dissociative Effect (EDE), and in green the Expected Associated Effect for the negative stratum (EAE$_-$).
\label{fig:application results}}
\end{figure}

Overall, we can conclude that the \PM exposure does have a negative effect on social mobility, coherently to the conclusion of \citet{Lee2024}. However, the effects are negative even for the dissociative strata (with exception of high school as post-treatment variable), meaning that even when \PM does not hinder educational attainments, the effects on social mobility are negative, reducing it. This last evidence hints at the possibility of further, yet unexplored, pathways connecting air pollution and social mobility. 

\subsection{Characterizing the Principal Strata}
\label{subsec:results}

Analyzing the distribution of characteristics between strata provides insights for designing targeted social policies to promote equity within the population. Figure \ref{fig:spiderplots} highlights that the primary factors that differ between strata include population density, median household income, and ethnic composition (percentage of white and black populations) and weather conditions.

%Weather conditions exhibit varying patterns based on education levels: counties with higher temperatures and more rainfall are associated with the positive associative stratum when high school is the  of the causal effect of exposure to \PM on social mobility. In contrast, for college and community college as mediators, these weather features define the negative associative stratum.

A consistent finding across the three levels of education is that counties with higher population density and higher median household income---typically urban areas---are more adversely affected by \PM exposure, leading to significant reductions in social mobility (negative associative stratum). Furthermore, counties with a higher percentage of white residents are predominantly assigned to the positive associative stratum, aligning with literature that indicates that white populations are less impacted by the adverse effects of air pollution and are also more likely to experience upward social mobility. In contrast, counties with a higher proportion of Black residents are concentrated in the negative associative stratum, emphasizing the possible need for targeted social policies to mitigate the increased risks faced by minority populations, particularly in urban areas, due to air pollution.

\begin{figure}
\begin{center}
\includegraphics[width=5.3in]{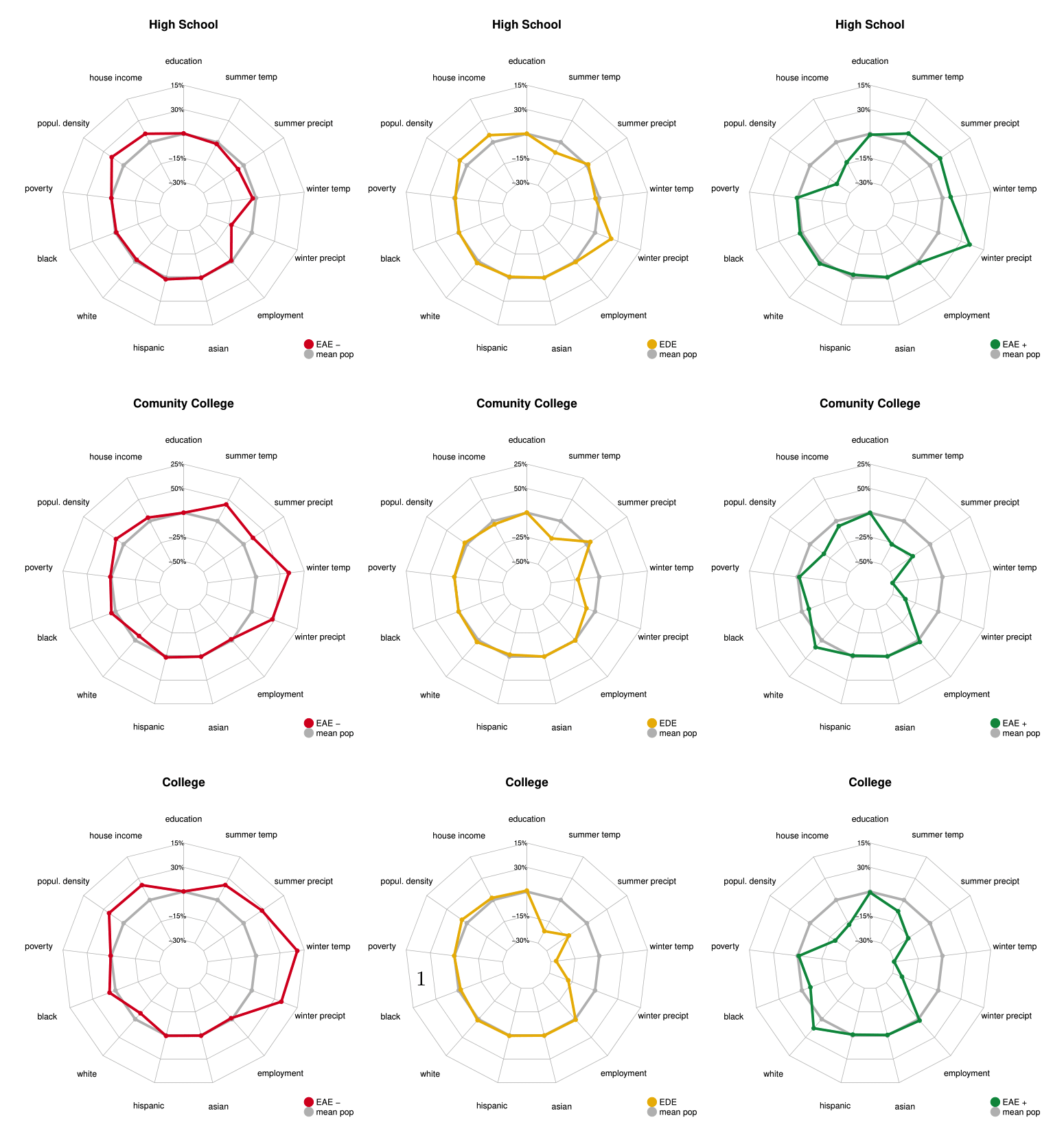}
\end{center}
\caption{Representation of the characteristics of the principal strata---$EAE_-$ in first column, $EDE$ in second column, and $EAE_+$ in the third column---for the three education level: high school (first row), community college (second row), and college (third row). Each spider plot reports in the colored area the strata-specific characteristics, the percentage of increment/decrement of the analyzed covariates compared with the mean among all the analyzed counties---reported in the gray lines.
\label{fig:spiderplots}}
\end{figure}

Moreover, we examine the spatial distribution of the identified principal strata, reported in Figure \ref{fig:maps_strata}. Some differences emerge in the central United States, particularly when comparing high school to community college and college strata. However, in all three levels of educational attainment, common spatial patterns can be observed. In particular, major cities tend to belong to the associative positive stratum  (shown in green on the maps), as evident in large metropolitan areas along the Northeast coast (Box A) and in California (Box B).
In contrast, certain regions characterized by high immigration rates, such as southern Florida (Box C) and areas along the Mexico border, particularly southern New Mexico (Box D), fall within the stratum where higher exposure to \PM is associated with lower educational attainment (associative negative stratum) and, consequently, a notable decline in social mobility.

\begin{figure}
\begin{center}
\includegraphics[trim={6cm 8cm 6cm 8cm},clip,width=5.5in]{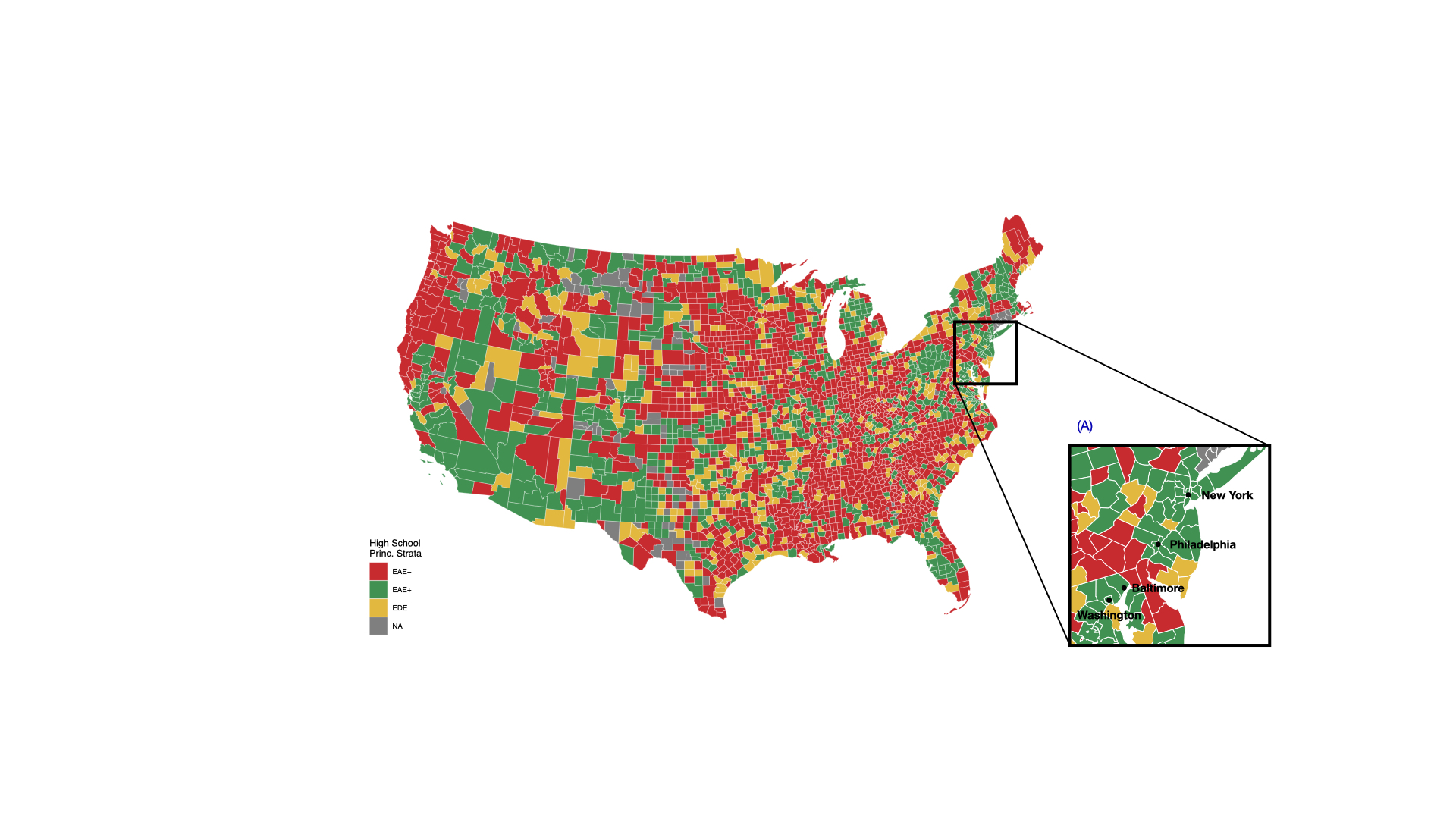}\\
\includegraphics[trim={6cm 5cm 6cm 8cm},clip,width=5.5in]{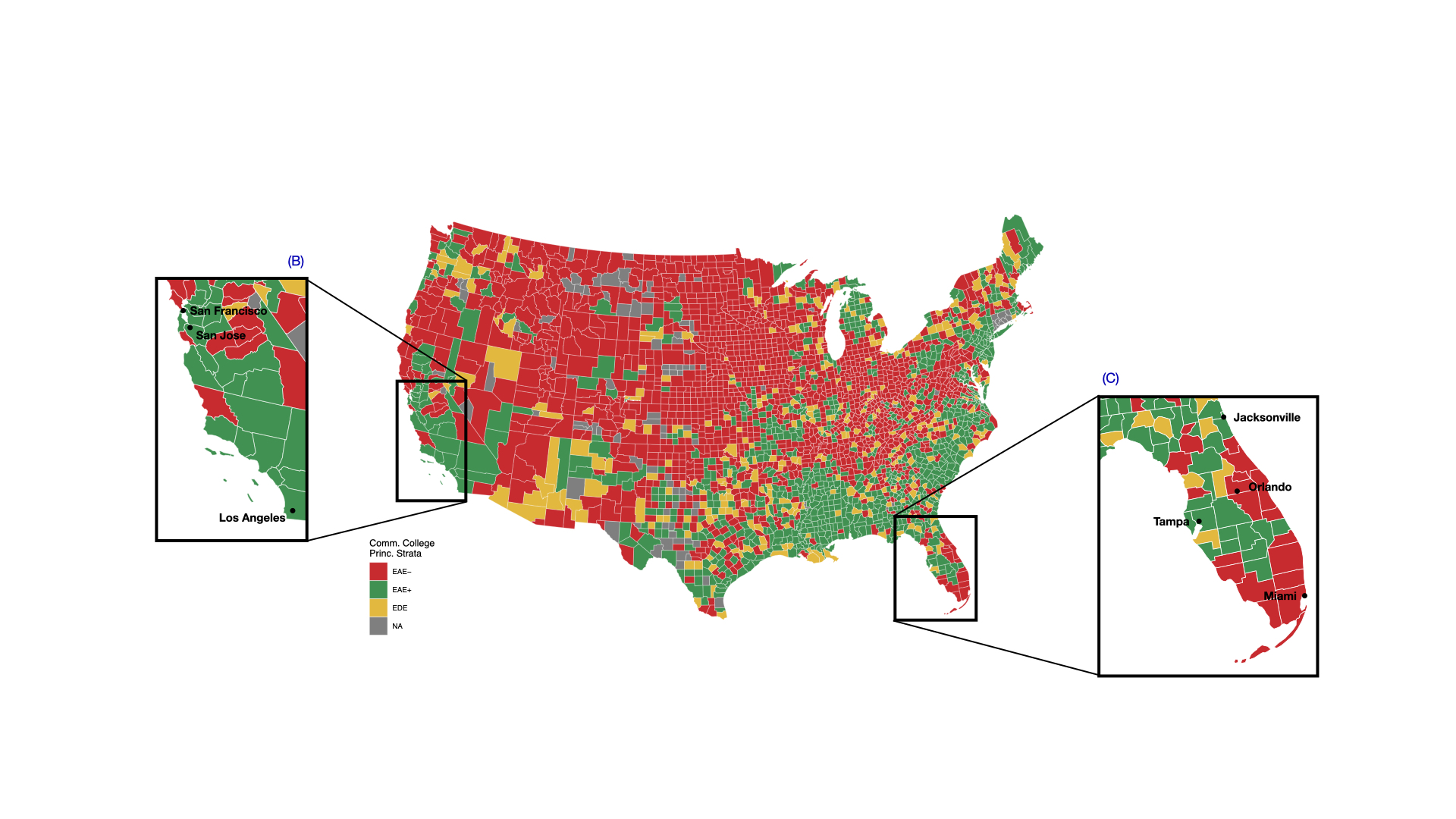}\\
\includegraphics[trim={6cm 2cm 6cm 10cm},clip,width=5.5in]{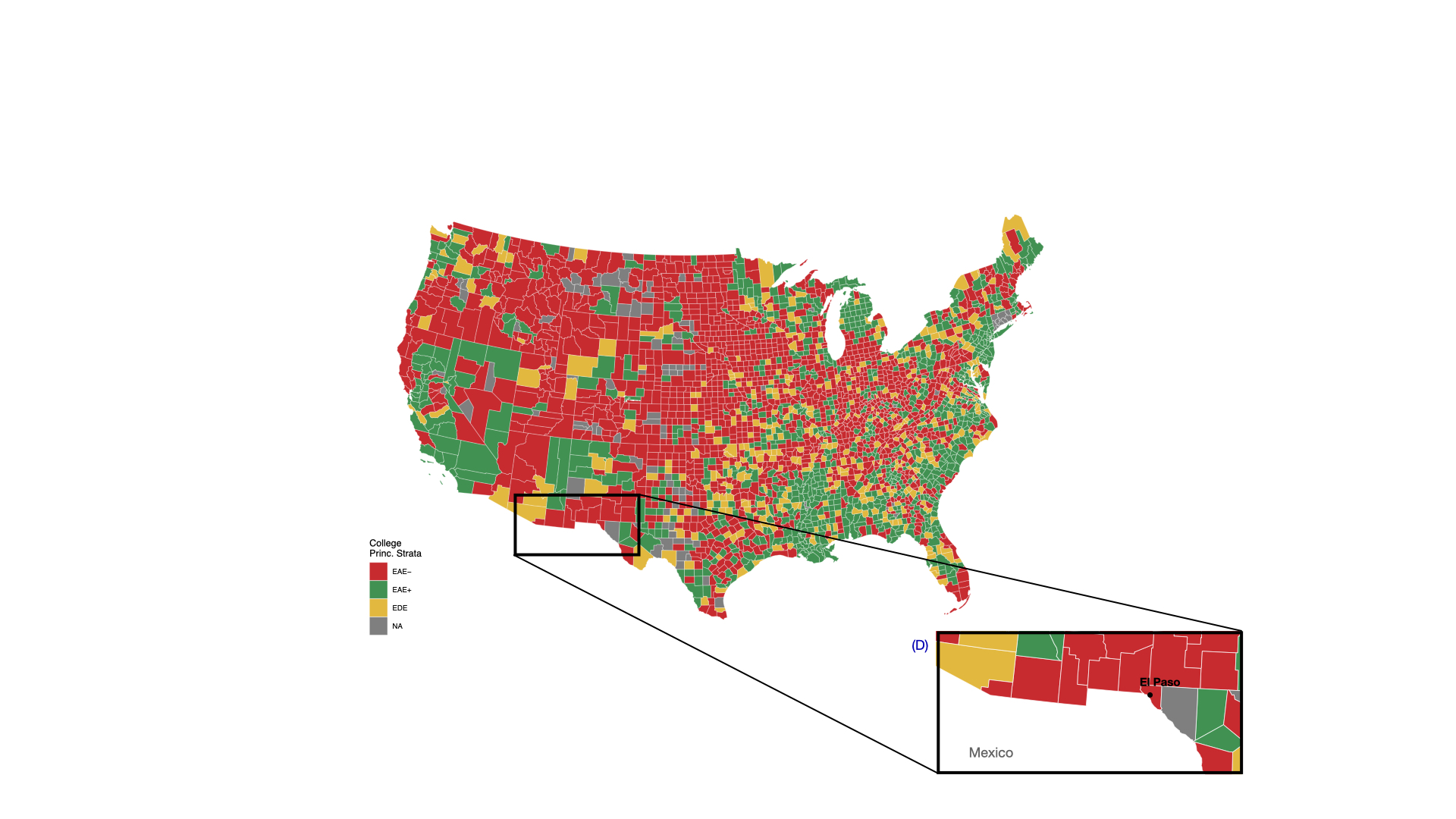}
\end{center}
\caption{Maps of the distributions of the identified principal strata for the three education levels: high school (top), community college (center), and college (bottom).
}\label{fig:maps_strata}
\end{figure}
\section{Conclusion}
\label{sec:conclusion}

In this paper, we proposed a novel Bayesian semi-parametric approach within the principal stratification framework. In particular, we proposed a DDP prior to flexibly modeling the distribution of potential primary outcomes. This approach induced a posterior predictive distribution for the post-treatment variable that was itself a mixture model, enabling accurate imputation of missing data for both the post-treatment variable and the primary outcome, which is crucial for accurately estimating principal causal effects.

The performance of the proposed model was evaluated in the simulation study, through different scenarios to test different levels of heterogeneity in the potential outcome for the post-treatment variable and for the primary outcome. The proposed model showed better performance than the model proposed by \citet{schwartz2011bayesian} and \citet{lu2023principal} in all scenarios.

In our motivating application, we disentangled the principal causal effects of air pollution on social mobility, through different levels of educational attainment served as a post-treatment variable. Specifically, we were interested in capturing the causal effects in three strata: counties where \PM exposure does not alter education levels and the two associative strata where education levels increase or decrease in expose to a higher level of \PMns. Overall, the results revealed a consistent negative effect of \PM exposure on social mobility at all three levels of education: high school, community college, and college rate. In counties where the higher \PM exposure either reduces or does not change education rates, the social mobility is significantly reduced.  This indicates that educational attainments are primary, but not the only causal pathway by which air pollution affects social mobility. Although in counties where the level of education increases with the higher level of \PM (a pattern observed in fewer than one-third of counties), no significant causal effect is measured on social mobility.

To our knowledge, this work is the first to study this complex causal link between air pollution and social mobility when education level serves as a post-treatment variable. Future research could investigate different post-treatment variables or consider simultaneously multiple ones.
Although our proposed method has been developed for our motivating application, it can be used for many other applications, where the causal pathway involves a similar variables definition.

From a methodological perspective, our approach could be extended by incorporating a nonparametric model for the post-treatment variable alongside the one used for the primary outcome, as suggested by \citet{schwartz2011bayesian}, \citet{zorzetto2024bayesian}, or \citet{antonelli2023principal}. In particular, adopting the nonparametric hierarchical prior introduced by \citet{zorzetto2024bayesian} would allow estimation of the principal strata without relying on predefined thresholds, enabling the data to guide the strata partitioning process. 
Another potential extension of our method involves accounting for spatial dependence by leveraging and adapting flexible models for causal inference, such as those proposed in \citet{duan2007generalized, petrone2009hybrid}. We leave these to future research.

\vspace{1cm}

\subsection*{Acknowledgments}
The authors wish to thank Luca Merlo and Sophie-An Kingsbury Lee for their helpful suggestions and comments for the application. 

%%%%%%%%%%%%%%%%%%%%%%%%%%%%%%%%%%%%%%%%%%%%%%
%% Funding information, if any,             %%
%% should be provided in the                %%
%% funding section.                         %%
%%%%%%%%%%%%%%%%%%%%%%%%%%%%%%%%%%%%%%%%%%%%%%
\subsection*{Funding}
This work was partially funded by the following grants: NIH:  R01MD016054, R01ES34021, R01ES037156, R01ES036436-01A1,  R01ES34021. This study was also supported by an AWS Research Project at Harvard T.H. Chan School of Public Health and UCLA funded by Amazon Web Services.

\subsection*{Code and Data} 
Code for implementing the proposed model and for replicating the results of simulation study and application is publicly available at \href{https://github.com/dafzorzetto/StrataBayes_SocialMobility}{\texttt{dafzorzetto/StrataBayes\_SocialMobility}}. The data are available at the original dataset, and the indication to obtain the analyzed dataset are reported at the same repository.
}

%%%%%%%%%%%%%%%

%-----------------------------------------------------------------------------------------------------
{\small
\bibliographystyle{chicago}
\bibliography{bibliography}
}
%-----------------------------------------------------------------------------------------------------
\pagebreak

\appendix
%\counterwithin{equation}{section}
%\counterwithin{figure}{section}
%\counterwithin{table}{section}
\pagenumbering{arabic}
\setcounter{page}{1}
    
\begin{center}
    \textbf{\large SUPPLEMENTARY MATERIAL TO\\ ``Characterizing the Effects of Environmental Exposures on Social Mobility: Bayesian Semi-parametrics for Principal Stratification''}\\ \vspace{0.25cm}
    \normalsize DAFNE ZORZETTO$^*$, PAOLO DALLA TORRE$^*$, SONIA PETRONE, FRANCESCA DOMINICI, AND FALCO J. BARGAGLI-STOFFI
\end{center}

\appendix
\counterwithin{equation}{section}
\counterwithin{figure}{section}
\counterwithin{table}{section}
\pagenumbering{arabic}
\setcounter{page}{1}

\section{Simulations details}
\label{app:simulation}

We report here the specific values used to simulate the three scenarios defined in Section 5.

\vspace{0.2cm}
\underline{\it Scenario 1}: The five confounders are simulated as: $X_1 \sim Be(0.4)$, $X_2 \sim Be(0.6)$, and the $X_3, X_4$ and  $X_5$ are sampled from a standard Gaussian distribution. The treatment variable is defined as $T \sim \mbox{Be}(\mbox{expit}(0.4X_{1}+0.4X_{2}+0.15X_{4}))$. The potential outcomes for the post-treatment variables distribution are defined as:

\begin{gather}
     P_i(0)\sim \Norm\Big(\beta^{(0)} {\bf X},\sigma_{p}^2\Big),\quad 
    P_i(1)\sim \Norm\Big(\beta^{(1)} {\bf X},\sigma_{p}^2\Big), \notag\
\end{gather}
and have the regression parameter reported in Table \ref{tab:simulation_1} with variance $\sigma_{p}^2$ equal to $1$ for both the treatment levels.

\begin{table}[ht]
    \caption{Scenario 1: values for the parameters for the potential post-treatment variables distribution.} 
        \begin{tabular}{l|cccccc}
             & Intercept & $X_1$ & $X_2$ & $X_3$ & $X_4$ & $X_5$  \\ \hline
            \( \beta^{(0)} \) & 1 & 2 & 3 & 0.5 & 0.1 & 0.3 \\ 
            \( \beta^{(1)} \) & 1 & 4 & 5 & 0.5 & 0.4 & 0.2 
        \end{tabular}
        \label{tab:simulation_1}
\end{table}

The three clusters are defined based on the two Bernoulli confounders $X_2$ and $X_3$. See Table \ref{tab:cluster_allocation} for the allocation criteria.

\begin{table}[ht]
    \centering
    \caption{Cluster allocation criteria based on confounders $X_2$ and $X_3$.}
    \begin{tabular}{l|cc}
        Cluster allocation & \( X_2 \) & \( X_3 \) \\ \hline
        $M=1$ & 1 & 1 \\ 
        $M=2$ & 1 & 0 \\ 
        $M=3$ & 0 & 0 \\ 
    \end{tabular}
    \label{tab:cluster_allocation}
\end{table}

The distributions for the two potential outcomes $\{Y(0),Y(1)\}$, conditional to the cluster allocation $M$ are defined as following:

\begin{gather*}
    Y(0)\mid M=m \sim \mathcal{N}\Big(\eta^{(0,m)}\Big[1,X_{1:3},-0.5|X_4|,e^{0.5X_5},P(0)\Big], {\sigma^{(0,m)}_{y}}^2\Big), \\
    Y(1) \mid M=m \sim \mathcal{N}\Big(\eta^{(1,m)}\Big[1,X_{1:3},-0.5|X_4|,e^{0.5X_5},P(1)-P(0)\Big], {\sigma^{(1,m)}_{y}}^2\Big),
\end{gather*}
where the parameter in the regression of the mean $\eta^{(t,m)}$ and the variance ${\sigma^{(t,m)}_{y}}^2$ depend on the allocated cluster $m$. In Table \ref{tab:etas_sett1} the values for the parameters are reported for each cluster.

\begin{table}[ht]
    \centering
    \caption{Scenario 1: values for the parameters for the potential outcomes distribution given the cluster allocation.}
        \begin{tabular}{l|cccccccc}
            Clusters & $\eta_1^{(0)}$ & $\eta_2^{(0)}$ & $\eta_3^{(0)}$ & $\eta_4^{(0)}$ & $\eta_5^{(0)}$ & $\eta_6^{(0)}$ & $\eta_7^{(0)}$ & ${\sigma^{(0)}_{y}}^2$ \\ \hline
            1 & 10 & 1.5 & 1.3 & 2 & 2 & 0.3 & 1.8 & 1\\ 
            2 & 1 & 1.1 & 0.75 & 1 & 0.2 & 0.3 & 0.2 & 2\\ 
            3 & -5 & 0.25 & 0.1 & 1 & 0.2 & 0.1 & -1 & 1.5\\ 
           \multicolumn{9}{c}{} \\
            Cluster & $\eta_1^{(1)}$ & $\eta_2^{(1)}$ & $\eta_3^{(1)}$ & $\eta_4^{(1)}$ & $\eta_5^{(1)}$ & $\eta_6^{(1)}$ & $\eta_7^{(1)}$ & ${\sigma^{(1)}_{y}}^2$\\ \hline
             1 & 3 & 1 & 1 & 1 & 0.3 & 0.3 & 1.5 & 2\\ 
            2 & 0.5 & 0.5 & 0.5 & 0.5 & 0.3 & 0.3 & 0.6 & 0.5 \\ 
            3 & -2 & 0.1 & 0.1 & 0.1 & 0.4 & 0.3 & -0.6 & 1
        \end{tabular}
        \label{tab:etas_sett1}
\end{table}

\vspace{0.2cm}
\underline{\it Scenario 2}: The ten confounders are simulated in the following way: $X_1 \sim Be(0.4)$ and $X_2 \sim Be(0.6)$---as in Scenario 1---, while the remaining eight confounders are sampled from a standard Gaussian distribution except for $X_4 \sim N(0,0.5)$ . The treatment level is defined as $T \sim \mbox{Be}(\mbox{expit}(0.2X_{1}+0.4X_{2}+0.1X_{6}))$. The cluster allocation is the same as that used in Setting 1 (see Table \ref{tab:cluster_allocation}). The potential post-treatment variables and potential outcomes distributions are defined as follows:
\begin{gather*}
    P(0)\sim \Norm\Big(\beta_2^{(0)} X_{3:10},1\Big),\quad 
    P(1)\sim \Norm\Big(\beta_2^{(1)} X_{3:10},1\Big),\\
    Y(0)\sim \sum_{m=1}^3\mathbb{I}_{\{M=m\}}\Big(\eta^{(0,m)}\Big[1,X_{3:6},|X_7+2|,e^{0.2X_{8}},-0.1|X_{9}|,e^{0.1X_{10}},P(0)],{\sigma^{(0,m)}_{y}}^2 \Big), \\
    Y(1)\sim \sum_{m=1}^3\mathbb{I}_{\{M=m\}}\Norm\Big(\eta^{(1,m)}\Big[1,X_{3:6},|X_7+1.5|,e^{0.1X_8},-0.3|X_{9}|,e^{0.2X_{10}},P(1)-P(0)\Big],\\
    {\sigma^{(1,m)}_{y}}^2\Big);\quad\quad\quad
\end{gather*}
where the values of the parameters involved are reported in Table \ref{tab:simulation_P_2} and in Table \ref{tab:etas_sett_2}, respectively for the potential outcome for the post-treatment distribution and the outcome distribution.

\begin{table}[ht]
    \centering
    \caption{Scenario 2: values for the parameters for the potential post-treatment variables distribution.}
        \begin{tabular}{l|ccccccccc}
            &Intercept & $X_3$ & $X_4$ & $X_5$ & $X_6$ & $X_7$ & $X_8$ & $X_9$ & $X_{10}$  \\ \hline
            \( \beta^{(0)} \) & -1 & 0.5 & 1.5 & 0.2 & 0.5 & 0.7 & 1 & -0.5 & -1.2  \\
            \( \beta^{(1)} \) & -0.5 & 1 & 1.8 & 0.2 & 0.5 & 0.7 & 1.2 & -0.3 & -1  
        \end{tabular}
        \label{tab:simulation_P_2}
\end{table}

\begin{table}[ht]
    \centering
    \caption{Scenario 2: values for the parameters for the potential outcomes distribution given the cluster allocation.}
        \begin{tabular}{l|cccccccccc}
            Cluster & intercept & $\eta_1^{(0)}$ & $\eta_2^{(0)}$ & $\eta_3^{(0)}$ & $\eta_4^{(0)}$ & $\eta_5^{(0)}$ & $\eta_6^{(0)}$ & $\eta_7^{(0)}$ & $\eta_8^{(0)}$ & \( {\sigma^{(0)}_{y}}^2 \)\\ \hline
            1 & 10 & 1.5 & 1.3 & 0.1 & 0.4 & 0.1 & 0.2 & -0.4 & 0.3 & 0.5\\ 
            2 & 2 & 1 & 1.1 & 0.75 & 0.1 & 0.1 & 0.2 & 0.2 & -0.4 & 0.5\\ 
            3 & -5 & 0.25 & 0.1 & 0.5 & 0.1 & 0.2 & 0.4 & -0.4 & -0.3 & 0.5\\ 
            \multicolumn{11}{c}{} \\
            Cluster & $\eta_1^{(1)}$ & $\eta_2^{(1)}$ & $\eta_3^{(1)}$ & $\eta_4^{(1)}$ & $\eta_5^{(1)}$ & $\eta_6^{(1)}$ & $\eta_7^{(1)}$ & $\eta_8^{(1)}$ & $\eta_9^{(1)}$ & \( {\sigma^{(1)}_{y}}^2 \)\\ \hline
            1 & 10 & 1 & 1 & 0.6 & 0.1 & 0.1 & 0.2 & -0.4 & 0.3 & 0.5\\ 
            2 & 2.5 & 0 & 0.8 & 0.5 & 0.5 & 0.1 & 0.2 & 0.4 & -0.4 & 0.5\\ 
            3 & -5 & 0.5 & 0.25 & 0.2 & 0.1 & 0.2 & 0.7 & -0.4 & 0.4 & 0.5
        \end{tabular}
        \label{tab:etas_sett_2}
\end{table}

\vspace{0.2cm}
\underline{\it Scenario 3}: The fourteen confounders are sampled in the following way: $X_1 \sim Be(0.4)$ and $X_2 \sim Be(0.6)$--as in scenario 1---, while the remaining twelve confounders are sampled from a Gaussian distribution with mean $0$ and variance that varies for each random variable taking values between $0.25$ and $1$. The treatment variable is sampled from a Bernoulli distribution $T \sim \mbox{Be}(\mbox{expit}(0.2X_{1}+0.4X_{2}+0.1X_{6}))$. The cluster allocation is the same as the one used in the first setting (see Table \ref{tab:cluster_allocation}). The potential outcomes for the post-treatment and outcome variable distributions are defined as:
\begin{gather*}
    P(0)\sim \Norm\Big(\beta_2^{(0)} X_{3:14},1\Big),\quad 
    P(1)\sim \Norm\Big(\beta_2^{(1)} X_{3:14},1\Big),\\
    Y(0)\sim \sum_{m=1}^3\mathbb{I}_{\{M=m\}}\Big(\eta^{(0,m)}\Big[1,X_{3:6},|X_7+2|,e^{0.2X_{8}},-0.1|X_{9}|,e^{0.1X_{10}},X_{11:14},P(0)],\\
    {\sigma^{(0,m)}_{y}}^2 \Big),\quad\quad\quad \\
    Y(1)\sim \sum_{m=1}^3\mathbb{I}_{\{M=m\}}\Norm\Big(\eta^{(1,m)}\Big[1,X_{3:6},|X_7+1.5|,e^{0.1X_8},-0.3X_{9},e^{0.2X_{10}},X_{11:14},\\
    P(1)-P(0)\Big],{\sigma^{(1,m)}_{y}}^2\Big);
\end{gather*}
where the values of the parameters involved are reported in Table \ref{tab:simulation_P_3} and in Table \ref{tab:etas_sett_3}, for the potential outcomes of the post-treatment and outcome variable distributions. The variances for the potential outcomes of the post-treatment and outcome variables are the same between each other, between treatment levels and across cluster allocations --- ${\sigma_p^{(t)}}^2 ={\sigma^{(t,m)}_{y}}^2=0.5$, for $t=0,1$ and for $m=\in\{1,2,3\}$. 

\begin{table}[ht]
    \centering
    \caption{Scenario 3: values for the parameters for the potential post-treatment variables distribution.} 
        \begin{tabular}{l|ccccccccccccc}
            & Intercept & $X_3$ & $X_4$ & $X_5$ & $X_6$ & $X_7$ & $X_8$ & $X_9$ & $X_{10}$ & $X_{11}$ & $X_{12}$ & $X_{13}$ & $X_{14}$ \\ \hline
            \( \beta^{(0)} \) & -1 & 0.5 & 1.5 & 0.2 & 0.5 & 0.7 & 1 & -0.5 & -1.2 & 0.1 & 0.1 & 0.1 & 0.1 \\ 
            \( \beta^{(1)} \) & -0.5 & 1 & 1.8 & 0.2 & 0.5 & 0.7 & 1.2 & -0.3 & -1 & 0.1 & 0.1 & 0.1 & 0.1 
        \end{tabular}
        \label{tab:simulation_P_3}
\end{table}

\begin{table}[ht]
    \caption{Scenario 3: values for the parameters for the potential outcomes distribution given the cluster allocation.}
        \begin{tabular}{l|cccccccccccccc}
            Clusters & \scriptsize $\eta_1^{(0)}$ & \scriptsize $\eta_2^{(0)}$ & \scriptsize $\eta_3^{(0)}$ & \scriptsize $\eta_4^{(0)}$ & \scriptsize $\eta_5^{(0)}$ & \scriptsize $\eta_6^{(0)}$ & \scriptsize $\eta_7^{(0)}$ & \scriptsize $\eta_8^{(0)}$ & \scriptsize $\eta_9^{(0)}$ & \scriptsize $\eta_{10}^{(0)}$ & \scriptsize $\eta_{11}^{(0)}$ & \scriptsize $\eta_{12}^{(0)}$ & \scriptsize $\eta_{13}^{(0)}$ & \scriptsize $\eta_{14}^{(0)}$  \\ \hline
            1 & 10 & 1.5 & 1.3 & 0.1 & 0.4 & 0.1 & 0.2 & -0.4 & 0.3 & 0.1 & 0.1 & 0.1 & 0.1 & 2 \\ 
            2 & 2 & 1 & 1.1 & 0.75 & 0.1 & 0.1 & 0.2 & 0.2 & -0.4 & 0.1 & 0.1 & 0.1 & 0.1 & 1 \\ 
            3 & -5 & 0.25 & 0.1 & 0.5 & 0.1 & 0.2 & 0.4 & -0.4 & -0.3 & 0.1 & 0.1 & 0.1 & 0.1 & -1 \\ 
            \multicolumn{15}{c}{} \\
            Clusters & \scriptsize $\eta_1^{(1)}$ & \scriptsize $\eta_2^{(1)}$ & \scriptsize $\eta_3^{(1)}$ & \scriptsize $\eta_4^{(1)}$ & \scriptsize $\eta_5^{(1)}$ & \scriptsize $\eta_6^{(1)}$ & \scriptsize $\eta_7^{(1)}$ & \scriptsize $\eta_8^{(1)}$ & \scriptsize $\eta_9^{(1)}$ & \scriptsize $\eta_{10}^{(1)}$ & \scriptsize $\eta_{11}^{(1)}$ & \scriptsize $\eta_{12}^{(1)}$ & \scriptsize $\eta_{13}^{(1)}$ & \scriptsize $\eta_{14}^{(1)}$\\ \hline
            1 & 10 & 1 & 1 & 0.6 & 0.1 & 0.1 & 0.2 & -0.4 & 0.3 & 0.1 & 0.1 & 0.1 & 0.1 & 2.5 \\ 
            2 & 2.5 & 0 & 0.8 & 0.5 & 0.5 & 0.1 & 0.2 & 0.4 & -0.4 & 0.1 & 0.1 & 0.1 & 0.1 & 0.5 \\ 
            3 & -5 & 0.5 & 0.25 & 0.2 & 0.1 & 0.2 & 0.7 & -0.4 & 0.4 & 0.1 & 0.1 & 0.1 & 0.1 & -2
        \end{tabular}
        \label{tab:etas_sett_3}
\end{table}
\newpage
\section{Posterior Inference}

In this section we describe the Gibbs sampler which allows us to draw from the posterior distribution. Following a number of iterations $r=1,\dots,R$ and using the observed data $(y,p,t,x)$ we were able to update parameters and impute the missing post-treatment $P^{mis}$ and both the observed and missing outcomes for the outcome $\{Y^{obs}, Y^{mis}\}$.

As we specified previously, the post-treatment variables are modeled with a linear regression and we leverage the conjugacy properties of the normal distribution to define the prior distribution of the regression parameters. We specify the prior distribution for the parameters involved in the post-treatment model:

\begin{gather*}
   \boldsymbol{\beta}^{(t)}\sim \Norm_{q+1}(\mu_\beta, \sigma^2_\beta \mathbf{I}_{q+1}),
    \\
    {\sigma_p^{(t)}}^2 \sim \mathcal{IG}(\gamma_1,\gamma_2).
\end{gather*}
Where  $\betat$ and $\sigmat_p^2$ are respectively the regression parameters of the mean and the variance of the post-treatment variable $P(t)$.

%in the following way:

%\begin{gather*}
%    \{\betat|P(t),\sigmat^2,\boldsymbol{X},\mu_p,\sigma_{\beta}^2)\} \sim \mathcal{N}(\mu_{\beta^{(t)}}^{new},\Sigma_{\beta^{(t)}}^{new}),
%    \\[10pt]
%    \Sigma_{\beta^{(t)}}^{new}=(\sigma_{\beta^{(t)}}^{-2}I + \boldsymbol{X}^T\boldsymbol{X}\sigmat^{-2})^{-1},
%    \\[10pt]
%\mu_{\beta^{(t)}}^{new}=\Sigma_{\beta^{(t)}}^{new}\times(\sigma_\beta^{(t)-2}I\mu_\beta + \boldsymbol{X}^T\boldsymbol{X}P(t)),
 %   \\[10pt]
  %  \{\sigmat^2|P(t),\beta^{(t)},\boldsymbol{X},\gamma_1,\gamma_2)\} \sim \mathcal{IG}\left(\gamma_1 + \frac{n_1}2,\gamma_2+\sum_{n=1}^{n_1}\frac{(P_i(t)-\beta^{(t)}\boldsymbol{X}{i})^2}{2}\right),
%\end{gather*}

%where $\sigma_{\beta^{(t)}}^{-2}$ represents the hyperparameter and  ${\sigma^{(t)}}^2$ represents the variance for the $\beta^{(t)}$ which is updated at each step of the Gibbs sampler.

%We model the outcome variable $Y(t)$, for each $t\in\{0,1\}$, with a Bayesian nonparametric approach, specifically we are using the Dependent Dirchlet Process(DDP) which allows us to learn information from the covariates in a flexible manner and it also identifies and estimates the heterogeneous groups.

Since we model the outcome variable $Y(t)$, for each $t\in\{0,1\}$, with a Bayesian mixture distribution, the cluster allocation $V_i(t)$, follows a Multinomial distribution with parameter $\lambda^{(t)}$:
\begin{equation*}
    V_i(t) \sim \mathcal{MN}(\boldsymbol{\lambda}^{(t)}(x_i)).
\end{equation*}

For computational reasons, we truncate the infinite mixture to $M$ components, such that
the vector of probabilities $\boldsymbol{\lambda}^{(t)}(x_i)$
\begin{gather*}
    \boldsymbol{\lambda}^{(t)}(x_{i}) = (\lambda^{(t,1)}(x_{i}), \dots, \lambda^{(t,M)}(x_{i})), \mbox{ for } t=\{0,1\}, %\\\Pr(V_{i}(t) = m) = \lambda^{(t,m)}(x_{i}),
\end{gather*}

Each probability $\lambda^{(t,m)}$ takes value in $[0,1]$, such that the $\sum_{m=1}^M \lambda^{(t,m)}=1$. The choice of $M$ can depend on each real-data scenario, according with the sample size. However, a large value is preferable.

As defined in the previous section, the outcome $Y(t)$ distribution, for each $t\in\{0,1\}$, given a cluster assignment follows a normal distribution. Specifically, we define the regression mean for the potential outcomes for the outcome $\{Y(0), Y(1)\}$ as:
\begin{gather}
   \{ Y_{i}(0)|\x_{i},p_i(0),p_i(1),\theta_y,V_{i}^{(0)}=m_0\} \sim {\mathcal N}\bigg(\boldsymbol{\eta}^{(0,m_0)}[1, \x_i,p_i(0)],{\sigma_y^{(0,m_0)}}^2\bigg),\notag \\
    \{ Y_{i}(1)|\x_{i},p_i(0),p_i(1),\theta_y,V_{i}^{(1)}=m_1\} \sim {\mathcal N}\bigg(\boldsymbol{\eta}^{(1,m_1)}[1, \x_i, p_i(1)-p_i(0)],{\sigma_y^{(1,m_1)}}^2\bigg),
\end{gather}

where:
\begin{gather*}
    \boldsymbol{\eta}^{(0,m_0)}[1, \x_i,p_i(0)]  =\eta_{0}^{(0,m)} + \eta_{1}^{(0,m)}X_i + \eta_{2}^{(0,m)}p_i(0),
    \\
    \boldsymbol{\eta}^{(1,m_1)}[1, \x_i, p_i(1)-p_i(0)] = \eta_{0}^{(1,m)}  +\eta_{1}^{(1,m)}X_i +\eta_{2}^{(1,m)}\big(p_i(1) - p_i(1)\big),
\end{gather*}

we will be using $X^{(y_0)} = [1, \x_i,p_i(0)]$ and $X^{(y_1)}=[1, \x_i, p_i(1)-p_i(0)]$ throughout the rest of the supplementary materials for conciseness.

%Note that while $X^{(y_0)}$ contains only information for $P(0)$, $X^{(y_1)}$ contains information for both $P(0)$ and $P(1)$. This choice is driven by the assumption that the outcome under control depend on the post-treatment variables under control, while we want to identify the effect of both the potential post-treatment variables on the outcome under treatment.

The prior distribution for the parameters used for the potential outcomes of the outcome $\{Y(0), Y(1)\}$ are the following:
\begin{equation*}
    \boldsymbol{\eta}^{(t,m)}\sim \mathcal{N}(\mu_\eta ,{ \sigma_{\eta}^{(t)}}^2 I), \quad \mbox{and} \quad
   {\sigma_y^{(t,m)}}^2\sim \mathcal{IG}(\gamma_1,\gamma_2).
\end{equation*}

The Gibbs sampler can be divided into five main parts.
\subsection{Imputation of missing post-treatment variable}
%\vspace{0.25cm}
%\subsubsection{}
%\noindent {\bf 1. Imputation of missing post-treatment variable:}
Due to missing data in the post-treatment variables, we define the post-treatment variable distribution that we use to sample from.
The post-treatment variable distributions differ according to the treatment level, due to different involvement of the post-treatment variables in the outcome model.

The post-treatment variable distribution under control is proportional to the distribution where this variable is involved and the data are observed. Therefore it is a mixture of normal distributions:

\begin{align*}
& \big\{P_i(0)|X_i,Y(1),\beta^{(0)},{\sigma_p^{(0)}}^2,\eta^{(0)},{\sigma_{y}^{(0)}}^2\big\}\\
& \propto \mathcal{N}(P_i(0);\beta^{(0)}X_i^p,{\sigma_p^{(0)}}^2)\sum_{m=1}^M\lambda^{(1,m)}(X_i)\mathcal{N}(Y_i(1);\eta^{(1,m)}X^{(y_0)}, {\sigma_{y}^{(1,m)}}^2) \\
    %P_i(0)|X_i,t,Y(1),\beta^{(0)},{\sigma^{(0)}}^2,\eta^{(0)},\sigma_{y^{(0)}} \sim \mathcal{N}(\beta^{(0)}X_i^p,{\sigma^{(0)}}^2)\sum_{m=1}^M\lambda^{(1,m)}(X_i)\mathcal{N}(\eta^{(1,m)}X^{(y_1)}, {\sigma_{y_1}^{(m)}}^2) \\
     &\propto \sum_{m=1}^M\lambda^{(1,m)}(X_i)\left[\mathcal{N}\Big(P_i(0);\beta^{(0)}X_i^p,{\sigma_p^{(0)}}^2\Big)\mathcal{N}\Big(Y_i(1);\eta^{(1,m)}X^{(y_0)}, {\sigma_{y}^{(1,m)}}^2\Big)\right]\\ 
     %&\propto \sum_{m=1}^M \lambda^{(1)}_m(x_i)Pr(P_i(0)|V_i(1)=m\ldots) 
     &= \sum_{m=1}^M \lambda^{(1,m)}(x_i)Pr\Big(P_i(0)|V_i(1)=m, X_i,t,Y(1),\beta^{(0)},{\sigma_p^{(0)}}^2,\eta^{(0)},{\sigma_{y}^{(1,m)}}^2\Big).
\end{align*}

The $Pr\Big(P_i(0)|V_i(1)=m, X_i,t,Y(1),\beta^{(0)},{\sigma_p^{(0)}}^2,\eta^{(0)},{\sigma_{y}^{(0)}}^2\Big)$ is the probability distribution of the post-treatment variable distribution under control given the allocation in the cluster $m$ for the outcome variable $Y_i(1)$. Such probability can be rewritten as:

\begin{align*}
    & \{P_i(0)|V_i(1)=m, X_i,t,Y(1),\beta^{(0)},{\sigma_p^{(0)}}^2,\eta^{(0)},{\sigma_{y}^{(0)}}^2\}\\
    & \quad\quad\quad \sim \mathcal{N}\left(\left(\frac{\beta^{(0)}X_i{\sigma_{y}^{(1)}}^2+{\sigma_p^{(0)}}^2\eta_{2}^{(1,m)}C_i^{(m)}}{{\sigma_{y}^{(1,m)}}^2+{\sigma_p^{(0)}}^2\eta_{2}^{(1,m)}}\right), \left(\frac{1}{{\sigma_p^{(0)}}^2} + \frac{{\eta_{2}^{(1,m)}}^2}{{\sigma_{y}^{(1,m)}}^2} \right)^{-1} \right),
\end{align*}

where
\begin{equation*}
    C_i^{(m)}=Y_i(1)-\eta_{0}^{(1,m)}-\eta_{1}^{(1,m)}X_i-\eta_{3}^{(1,m)}P_i(1).
\end{equation*}

\vspace{0.25cm}
\subsection{Cluster Allocation}
%\noindent {\bf 2. Cluster allocation:} 
The cluster allocation is driven by the categorical latent variable $V_i(t)$ for each unit $i\in\{1,\dots, n\}$ and for each treatment level $t\in\{0,1\}$. This variable has multivariate distribution:
\begin{equation*}
    V_i(t) \sim \mathcal{MN}\left(\boldsymbol{\lambda}^{(t)new}(X_i)\right),
\end{equation*}
where:
\begin{gather*}
    \boldsymbol{\lambda}^{(t)new}(X_i) = \left(\lambda^{(t,1)new}(X_i) \dots \lambda^{(t,M)new}(X_i) \right) , \\
    \lambda^{(t,m)new}= Pr(V_i(t)=m) \propto \lambda^{(t,m)}(X_i)\mathcal{N}\Big(y^{(t)}; \eta^{(t,m)}X^{(y_t,m)}, {\sigma_{y}^{(t,m)}}^2\Big) ,
\end{gather*}

\vspace{0.25cm}
\subsection{Augmentation scheme}
%\noindent {\bf Augmentation scheme:} 
For computational efficiency, we leverage an augmentation scheme to estimate the posterior distribution of the regression parameters in the weights of the mixture.
Therefore, we have to introduce the latent variable $Z_i^{(t)}(X_i)$, for each unit $i\in\{1,\dots, n\}$ and for each treatment level $t\in\{0,1\}$ such that 

%\begin{equation*}
%    \left\{ Z_i^t(X_i)|V_i(t)=m, \gamma_m^t(X_i) \right\},
%\end{equation*}

\begin{equation*}
\left\{ Z_i^{(t)}(X_i) \mid V_i(t)=m, \gamma^{(t,m)}(X_i) \right\} \sim \left\{
\begin{aligned}
  &\mathcal{N}(\gamma^{(t,m)}(x_i), 1) \mathbbm{1}_{\mathbb{R}^+} &\text{if } V_i(t)=m, \\
  &\mathcal{N}(\gamma^{(t,m)}(x_i), 1) \mathbbm{1}_{\mathbb{R}^-} &\text{if } V_i(t) < m.
\end{aligned}
\right.,
\end{equation*}
Where the parameters $\{\gamma^{(t,m)}(X_i)\}_{m=1}^{M}$ are computed recursively:
\[ \begin{aligned}
    &\gamma^{(t,1)}(X_i) = \Phi^{-1} \left(\lambda^{(1)}(X_i) \right)\\
    &\vdots\\
    &\gamma^{(t,m)}(X_i) =  \Phi^{-1}\left(\frac{\lambda^{(t,m)}(X_i)}{1- \sum_{a<m}\lambda^{(t,a)}(X_i)} \right)
\end{aligned},
\]

Therefore, we obtain the matrices $\Tilde{Z^{(t)}}$ such that
\begin{gather*}
    \Tilde{Z^{(t)}} \sim \mathcal{N} \left(\varepsilon_{0n}^{(t)} + \Tilde{X}^T\varepsilon_n^{(t)}, 1  \right), \\
    \varepsilon_n^{(t)}=(\varepsilon_{0n}^{(t)},\varepsilon_n^{(t)})^T \sim \mathcal{N}_{p+1}\left(\mu_{\varepsilon},\sigma_{\varepsilon}^2I\right),
\end{gather*}
that allows us to write the posterior distribution for the regression parameters in the weights as:
\begin{equation*}
    \varepsilon_m^{(t)}|\Tilde{Z} \sim \mathcal{N}_{p+1}\left(v^{-1}n,v^{-1}\right),
\end{equation*}

where $v=\left(\frac{1}{\sigma_{\varepsilon}^2}I + \frac{\Tilde{X}^T\Tilde{X}}{1} \right)$ and $n= \frac{\mu_\varepsilon}{\sigma_{\varepsilon}^{2}}I + \Tilde{X}^T\Tilde{Z}$.

\vspace{0.25cm}
\subsection{Cluster-specific parameters}
%\noindent {\bf 4. Cluster-specific parameters:} 
The posterior distribution for the cluster-specific parameters under control are the following:

\begin{align*}
    \{\eta_0^{(m)}|Y_i(0)...\} &\sim \mathcal{N}_{2+p}\left(\mu_{\eta}^{(m)new}, \Sigma_{\eta}^{(m)new} \right), \\
    {\{\sigma_{y^{(0)}}^{(m)}}^2|Y_i(0)\dots\} &\sim \mathcal{IG}\left(\gamma_{y_1} + \frac{n^{(0,m)}}{2} \ , \ \gamma_{y_2} + \sum_{n=1}^{n^{(0,m)}}\frac{(Y_i(0)-\eta^{(0,m)}X^{(y_0,m)})^2}{2}\right).
\end{align*}

In particular, we have that 
\begin{align*}
    \Sigma_{\eta}^{(m)new} &= \left(\sigma_{y^{(0)}}^{(m)^{-2}}X^{(y_0,m)^T}X^{(y_0,m)}+\sigma^{-2}_\eta I \right)^{-1}, \\
    \mu_{\eta}^{(m)new} &= \Sigma_{\eta}^{(m)new}\left(\mu_{\eta} \lambda_{\eta}^{-2}I + X^{(y_0,m)}Y(0)^{(m)} \right),
\end{align*}
where we indicate as $X^{(y_0,m)}= X^{(y_0)}_{\{row\ i\ :\ V_i(0)=m\}}$ and $Y(0)^{(m)} =Y(0)_{\{row\ i \ : \ V_i(0)=m\}}$, respectively the matrix of units allocated in the cluster $m$.

In similar way, the posterior distribution for the cluster-specific parameters under treatment are 
\begin{align*}
    \{\eta^{(1,m)}|Y_i(1) \dots \} &\sim \mathcal{N}_{3+p} \left( \mu_{\eta}^{(m)new}, \Sigma_\eta^{(m)new} \right)\\
    {\{\sigma_{y}^{(1,m)}}^2|Y_i(1)\dots\} &\sim \mathcal{IG}\left(\gamma_{y_1} + \frac{n_1^{(m)}}{2},\gamma_{y_2} + \sum_{n=1}^{n_1^{(m)}}\frac{(Y_i(1)-\eta^{(1,m)}X^{(y_1,m)})^2}{2}\right)
\end{align*}

where
\begin{align*}
\Sigma_{\eta}^{(m)new} &=\left(X^{(y_1,m)^T}\sigma_{y}^{(1,m)^{-2}}X^{(y_1,m)} +\lambda^{-2}_{\eta}I \right)^{-1}, \\
    \mu_{\eta}^{(m)new} &=\Sigma_{\eta}^{(m)new}\left(\mu_{\eta}\sigma_{\eta}^{-2}I + X^{(y_1,m)}Y(1)^{(m)} \right),
\end{align*}
with $X^{(y_1,m)} = \left[ 1 \quad X \quad P(0) \quad P(1) \right]_{\{row \ i: \ V_i(1)=m\}}$
 and $Y(1)^{(m)} =Y(1)_{\{row\ i \ : \ V_i(1)=m\}}$.

\vspace{0.25cm}
\subsection{Imputation of potential outcome}
%\noindent {\bf 5. Imputation of potential outcome:}
The distributions for the potential outcome- for each unit $i \in \{1, \dots, n\}$---under control $Y_i(0)$ and under treatment $Y_i(1)$---, given the allocation at the cluster $m_0$ and $m_1$ respectively, are the following:
\begin{align*}
    \{Y_i(0)|V_i(0)=m_0, \eta^{(0)}, {\sigma_{y}^{(0)}}^2\} &\sim \mathcal{N}\left( \eta^{(0,m)}X^{(y_0,m)}, {\sigma_{y}^{(0,m)}}^2 \right),\\
    \{Y_i(1)| V_i(1)=m_1, \eta_1, {\sigma_{y}^{(1)}}^2\} &\sim \mathcal{N}\left(\eta^{(1,m)}X^{(y_1,m)},{\sigma_{y}^{(1,m)}}^2\right).
\end{align*}
where the parameters involved have the posterior distributions defined in the previous step.
%distributions are the priors already defined: $\eta^{(0,m)}\sim \mathcal{N}_{2+p}\left( \mu_{\eta}, \lambda_{\eta}^2I\right)$ and ${\sigma_{y^{(0)}}^{(m)}}^2 \sim \mathcal{IG}(\gamma_{y_1},\gamma_{y_2})$.

%%%%%%%%%%%%%%%%%%%%%%%%%%%%%%%%%%%%%%%%%%%%%%%%%%%%%%%%%%%%%%

%\begin{comment}

%\subsubsection*{Cluster allocation and augmentation scheme}

%\begin{align*}
%    Pr(V_i(t)=m) = \lambda^{(t,m)}(X_i) = U^{(t,m)}(X_i)\prod_{a<m}\left(1 - U^{(t,a)}(X_i) \right) = \\ = \Phi\left(\gamma^{(t,m)}(X_i)\right)\prod_{a<m}\left(1 - \Phi(\gamma^{(t,a)}(X_i) \right) ,
%\end{align*}

%\begin{gather*}
%    \gamma^{(t,m)}(X_i) \sim \mathcal{N}(\varepsilon_{0}^{(t,m)} + X_i^T\varepsilon^{(t,m)}, 1) ,\\
%    V_i(t) \sim \mathcal{MN}\left(\boldsymbol{\lambda}^{(t)}(X_i)\right),
%\end{gather*}
%\end{comment}

\newpage
\section{More Applications Details}
\label{app:simulation}

In this section, we illustrate the distribution of the observed variable used in the real-data applications. Figure \ref{fig:maps_mobility} reports the absolute upward mobility \citep[AUM, ][]{chetty2017fading}, which is defined as the mean income percentile in adulthood of individuals born between 1978 and 1983 in families in the $25^{th}$ percentile of the national parent income distribution. Income rank is measured in 2015 (ages 31–37).

\begin{figure}
\begin{center}
\includegraphics[width=4.5in]{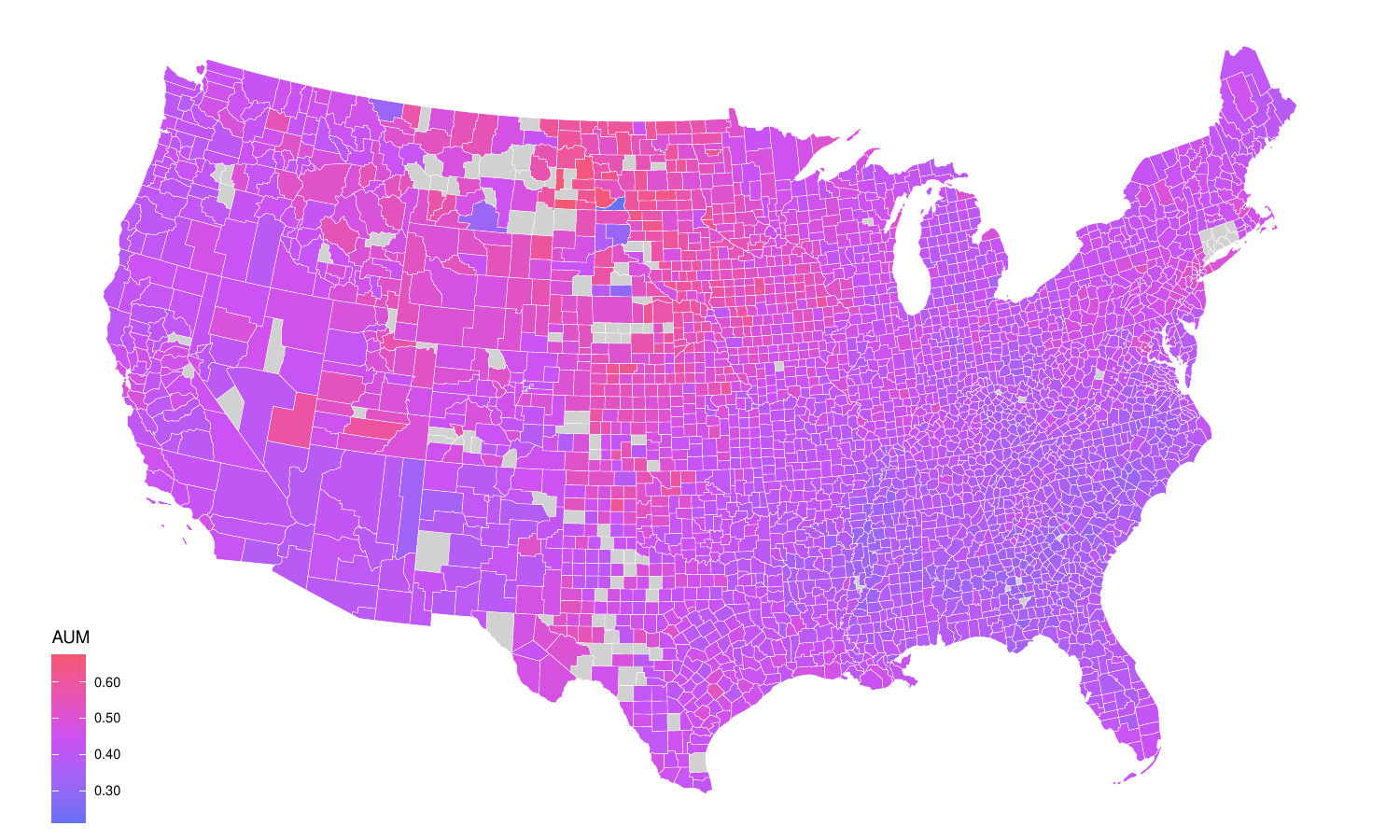}
\end{center}
\caption{Maps of the observed distributions for social mobility (AUM).
}\label{fig:maps_mobility}
\end{figure}

Figure \ref{fig:maps_schools} illustrates the observed distribution of the three categories to define educational attainments: community college, high school, and college.

\begin{figure}
\begin{center}
\includegraphics[width=4.2in]{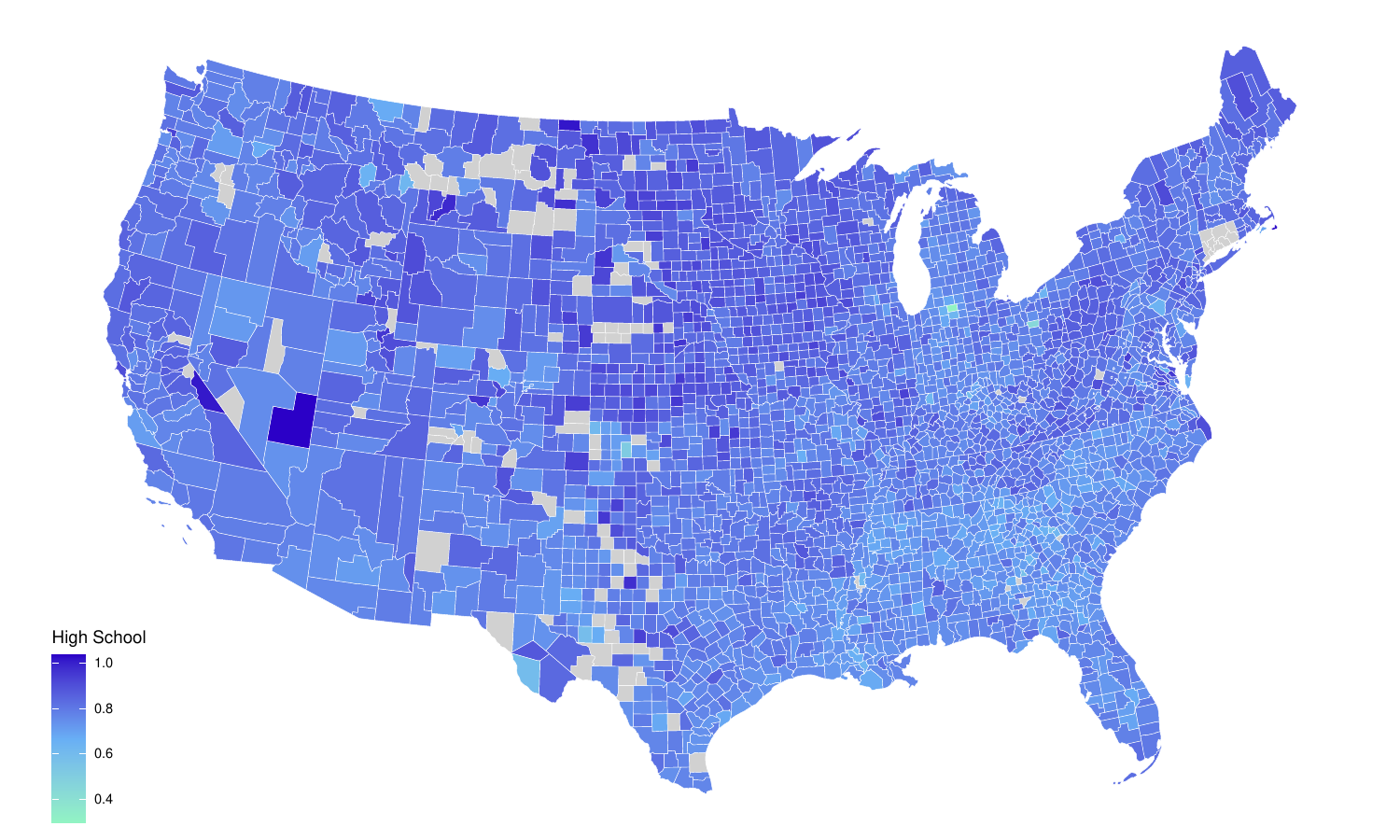}\\
\includegraphics[width=4.2in]{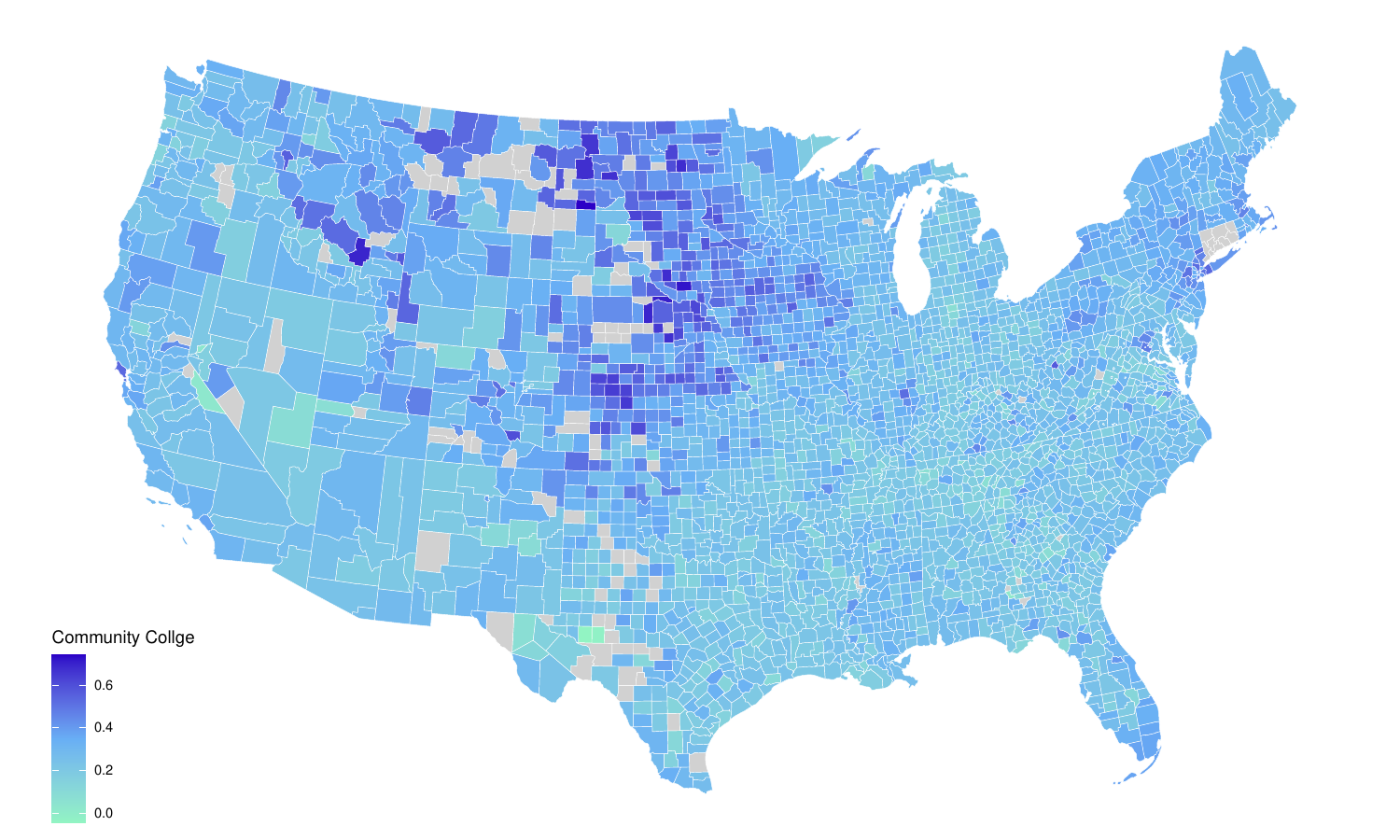}\\
\includegraphics[width=4.2in]{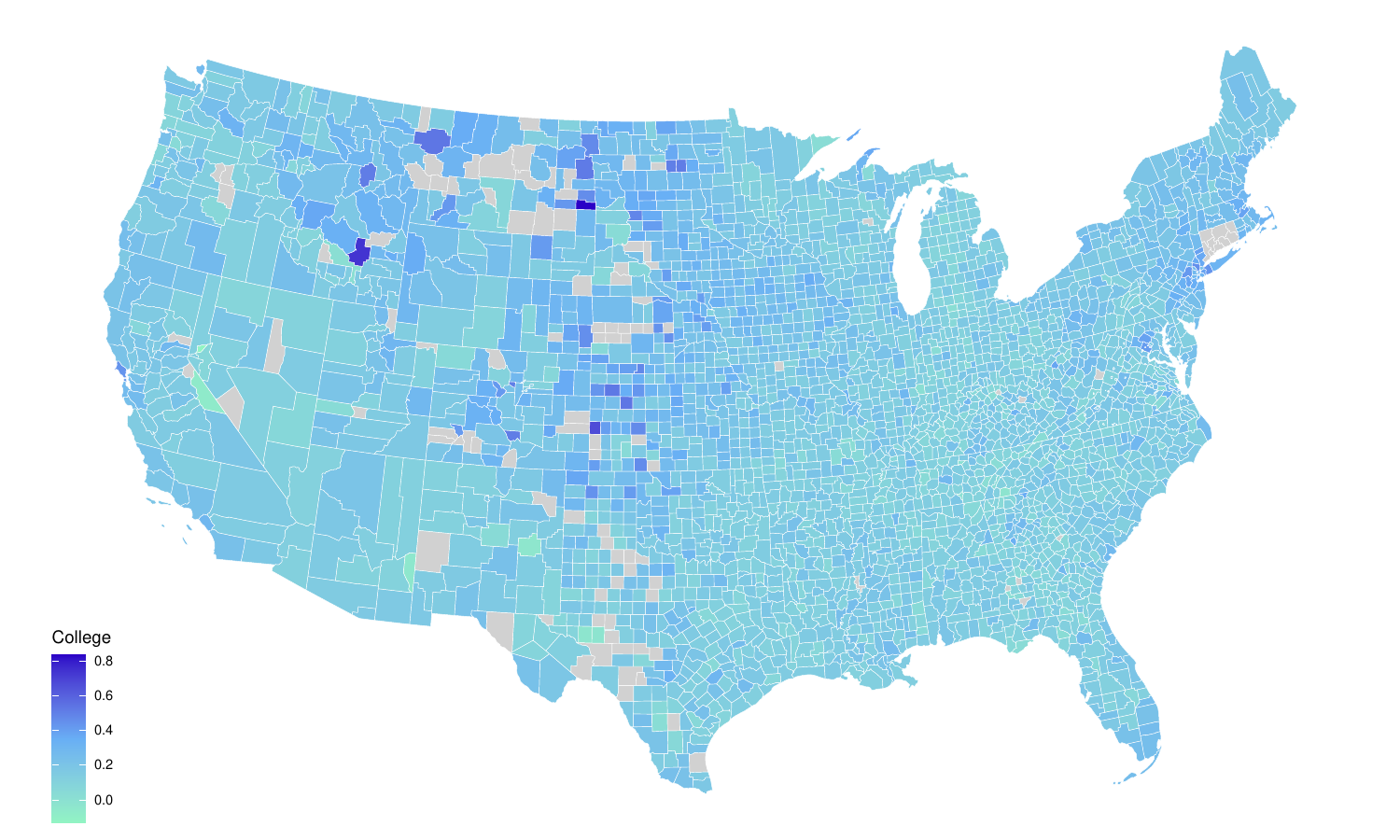}
\end{center}
\caption{Maps of the distributions of observed variables: high school attainment rate, community college attainment rate, and college attainment rate (in order from top to bottom).
}\label{fig:maps_schools}
\end{figure}

\end{document}